\documentclass[aps,prd,twocolumn,superscriptaddress,floatfix,nofootinbib]{revtex4}

\usepackage{color}
\usepackage[dvips]{graphicx}

\newcommand{\lsim}{\mathrel{\mathop{\kern 0pt \rlap
      {\raise.2ex\hbox{$<$}}}\lower.9ex\hbox{\kern-.190em $ \sim$}}}

\newcommand{\gsim}{\mathrel{\mathop{\kern 0pt
      \rlap{\raise.2ex\hbox{$>$}}}\lower.9ex\hbox{\kern-.190em $\sim$}}}

\newcommand{\beq}{\begin{equation}}
\newcommand{\eeq}{\end{equation}}

\newcommand{\be}{\begin{equation}}
\newcommand{\ee}{\end{equation}}
\newcommand{\beqarr}{\begin{eqnarray}}
\newcommand{\eeqarr}{\end{eqnarray}}
\newcommand{\sigmav}{\langle \sigma_{\rm ann} v \rangle}

\newcommand{\sigmavint}{\langle \sigma_{\rm ann} v \rangle_{\rm int}}

\begin{document}

\title{Embedding the 125 GeV Higgs boson measured at the LHC in an effective
MSSM: possible implications for neutralino dark matter}

%\thanks{Preprint number: DFTT 34/2011}


%
\author{S. Scopel}
\affiliation{Department of Physics, Sogang University\\
Seoul, Korea, 121-742}
\author{N. Fornengo}
\affiliation{Dipartimento di Fisica, Universit\`a di Torino \\
and Istituto Nazionale di Fisica Nucleare, Sezione di Torino \\
via P. Giuria 1, I--10125 Torino, Italy}
\author{A. Bottino}
\affiliation{Dipartimento di Fisica, Universit\`a di Torino \\
and Istituto Nazionale di Fisica Nucleare, Sezione di Torino \\
via P. Giuria 1, I--10125 Torino, Italy}

\date{\today}

\begin{abstract}

We analyze the phenomenological consequences of assuming that
the  125 GeV  boson measured at the LHC coincides with one of the
two CP--even Higgs bosons of an effective Minimal Supersymmetric
extension of the Standard Model  at the electroweak scale.
We consider the two ensuing scenarios and discuss critically
the role of the various experimental data (mainly obtained at colliders
and at B--factories) which provide actual or potential constraints
to supersymmetric properties. Within these scenarios, properties of
neutralinos as dark matter particles are analyzed from the point of view of
their cosmological abundance and rates for direct and indirect detections.

\end{abstract}


\maketitle

%%%%%%%%%%%%%%%%%%%%%%%%%%%%%%%%%%%%%%%%%%%%%%%%%%%%%%%%%%%%%%%%%%%%%%%%%%%%%
\section{Introduction}
\label{sec:intro}

It is remarkable that the neutral boson with a mass of
125--126 GeV, measured at the LHC in the diphoton, $Z Z$, $W W$ and
$\tau \tau$ channels
(hereafter denoted by $H_{125}$) \cite{lhc},
can be interpreted as the Higgs particle of the Standard Model (SM)  .
However, due to the well known problems of quadratic divergences related to the
Higgs mass, a pressing question is whether this newly discovered Higgs--like particle
can be interpreted within a supersymmetric extension of the SM, where the
problem of divergences would be solved by boson--fermion loop cancellations.
Should this be the case, a very rich and intriguing phenomenology would open up
\cite{hall,bbm,hsw,arbey,draper,carena1,phenomenology,chs,carena2,ben,drees,b,carena3,carena4,han}.

Here we investigate this possibility in detail, also in connection
with possible implications for supersymmetric candidates of dark
matter (DM) in the Universe.  We employ a simple supersymmetric model,
which we already used in previous analyses
\cite{lowneu,interpreting,discussing,phenomenology}, consisting in an
effective Minimal Supersymmetric extension of the Standard Model
(MSSM) at the electroweak (EW) scale, where the usual hypothesis of
gaugino--mass unification at the scale of Grand Unification (GUT) of a
supergravity (SUGRA) model, is removed; this effective MSSM is very
manageable, since expressible in terms of a limited number of
independent parameters.

The Higgs sector of this MSSM has two Higgs doublets, which generate,
by spontaneous symmetry breaking, two vev's: $v_1$ and $v_2$. These
provide masses to the down--type quarks and the up--type quarks,
respectively. As usual, an angle $\beta$ is introduced and defined as
$\tan \beta = v_2/v_1$. This Higgs sector contains three neutral
bosons: two CP--even, $h,H$, and a CP-odd one $A$.  The two CP--even
Higgs bosons are defined, in terms of the neutral components of the
original Higgs doublets, as: $H = \cos \alpha H^0_1 + \sin \alpha
H^0_2$, $h = - \sin \alpha H^0_1 + \cos \alpha H^0_2$. In the
diagonalization of their mass matrix, the mass hierarchy $m_h < m_H$
is imposed, and the angle $\alpha$ is taken in the range
$[-\pi/2,\pi/2]$.

The lower bound on $m_h$ can be established using the LEP data on the
search for Higgs particles \cite{lep}.  Contrary to the usual
assumption (employed in most of the literature until very recently) of
assuming for this lower limit the {\it Standard Model} bound $m_h >$
114 GeV, in our previous works (and in the present one) we take the
actual LEP constraint on the Higgs--production cross sections, that
can be translated into a bound on the quantity $\sin^2 (\alpha -
\beta)$ as a function of $m_h$ (this quantity represents the ratio of
the cross section for the Higgs--strahlung process $e^+ e^-
\rightarrow Z h$ to the corresponding SM cross section; a
complementary bound arises from the process $e^+ e^- \rightarrow Z A$,
and it is also taken into account).  In
Refs. \cite{lowneu,interpreting,discussing,phenomenology} we showed
that the LEP limit actually allows the mass $m_A$ to reach values as
low as 90 GeV in some regions of the supersymmetric parameter space in
MSSM, where the phenomenology of neutralino DM is particularly
interesting.

The ATLAS and CMS data exclude that the boson $A$ can be identified
with the new particle $H_{125}$, but we are left with the two
options: either $H \equiv H_{125}$ (hereafter denoted as scenario I)
or $h \equiv H_{125}$ (scenario II).

As pointed out in Ref. \cite{phenomenology}, scenario I arises
naturally in the supersymmetric scheme considered in Refs.
\cite{lowneu,interpreting,discussing}, when $m_h$ is taken as light as
possible (compatibly with the mentioned LEP bound). In fact, in this
regime one has $m_h \sim m_A \simeq (90 - 100)$ GeV, and $m_H \simeq
(115 - 130)$ GeV \cite{phenomenology}. This scenario has also been
discussed in Refs. \cite{hsw,b,chs,drees,carena4}.  As remarked in
Ref. \cite{drees,Bottino:2000kq}, the light $h$ boson of this scenario
could be the origin of the small excess of Higgs--like events observed
at LEP \cite{lep2003}.

The second option is represented by scenario II: this scenario occurs,
when the Higgs--like boson observed at LHC is identified with the
lighter CP--even boson $h$ within the MSSM
\cite{hall,bbm,hsw,arbey,draper,carena1,chs,carena2,ben,b,carena3,carena4,han}. In
this case, $m_h \simeq (125 - 126)$ GeV and $A$, $H$ can also decouple
substantially from $h$, but with $m_A \simeq m_H$.

In this paper we analyse separately scenarios I and II,
critically discussing  the role of the various experimental data
(mainly obtained at colliders
and at B--factories) which provide actual or potential constraints
to supersymmetric properties. For each scenario the  properties of
neutralinos as DM particles are then analyzed from the point of view of
their cosmological abundance and rates for direct and indirect detections.

The scheme of the presentation is the following. In
Sect. \ref{sec:model} the features of the employed MSSM are described,
in Sect. \ref{sec:constraints} a full list of conceivable constraints
on the model is introduced. Results are given in
Sects. \ref{sec:results},\ref{sec:direct},\ref{sec:indirect} and
conclusions in Sect. \ref{sec:conclusions}.

\section{Effective MSSM}
\label{sec:model}

The supersymmetric model we consider here is an effective MSSM scheme
at the electroweak scale, with the
following independent parameters: $M_1$, $M_2$, $M_3$, $\mu$,
$\tan\beta$, $m_A$, $m_{\tilde q_{12}}$, $m_{\tilde t}$, $m_{\tilde
  l_{12,L}}$, $m_{\tilde l_{12,R}}$, $m_{\tilde {\tau}_L}$, $m_{\tilde
  {\tau}_R}$ and $A$. We stress that the parameters are defined at the
EW scale.  Notations are as follows: $M_1$, $M_2$ and $M_3$ are the
U(1), SU(2) and SU(3) gaugino masses (these parameters are taken here
to be positive), $\mu$ is the Higgs mixing mass parameter, $\tan\beta$
the ratio of the two Higgs v.e.v.'s, $m_A$ the mass of the CP--odd
neutral Higgs boson, $m_{\tilde q_{12}}$ is a squark soft--mass common
to the squarks of the first two families, $m_{\tilde t}$ is the squark
soft--mass for the third family, $m_{\tilde l_{12,L}}$ and $m_{\tilde
  l_{12,R}}$ are the slepton soft--masses common to the L,R components
of the sleptons of the first two families, $m_{\tilde {\tau}_L}$ and
$m_{\tilde {\tau}_R}$ are the slepton soft--masses of the L,R components
of the slepton of the third family, $A$ is a common dimensionless
trilinear parameter for the third family, $A_{\tilde b} = A_{\tilde t}
\equiv A m_{\tilde t}$ and $A_{\tilde \tau} \equiv A (m_{\tilde
  {\tau}_L}+m_{\tilde {\tau}_R})/2$ (the trilinear parameters for the
other families being set equal to zero).  In our model, no gaugino
mass unification at a Grand Unified scale is assumed, and therefore
$M_1$ can be sizeably lighter than $M_2$.  Notice that the present
version of this framework represents an extension of the model discussed in
our previous papers \cite{lowneu,interpreting,discussing}, where a
common squark and the slepton soft mass was employed for the 3
families.

The linear superposition of bino $\tilde B$, wino $\tilde W^{(3)}$
and of the two Higgsino states $\tilde H_1^{\circ}$, $\tilde
H_2^{\circ}$ which defines the neutralino state of lowest mass $m_{\chi}$ is written here as:
\begin{equation}
\chi \equiv a_1 \tilde B + a_2 \tilde W^{(3)} +
a_3 \tilde H_1^{\circ} + a_4  \tilde H_2^{\circ}.
\label{neutralino}
\end{equation}
We assume R--parity conservation to guarantee that the lightest supersymmetric particle is stable (we consider only models where this is the neutralino).

Within our model we calculate all the quantities necessary to impose
the constraints discussed in Sect.  \ref{sec:constraints}, and the
cross sections relevant for direct and indirect detection of DM
neutralinos: the neutralino-nucleon cross section $\sigma_{\rm
  scalar}^{(\rm nucleon)}$ and the thermally averaged product of the
neutralino pair annihilation cross section times the relative velocity
$\sigmav$. 

The neutralino--nucleon scattering takes contributions from ($h, A,
H$) Higgs boson exchange in the t--channel and from the squark
exchange in the s-channel; the $A$--exchange contribution is
suppressed by kinematic effects. This cross section is evaluated here
according to the formulae given in Ref. \cite{indication}. For the
crucial coupling parameter $g_d$ entering the Higgs boson exchange
amplitude, we take its {\it reference} value $g_{d,\rm ref} = 290$ MeV
employed in our previous papers \cite{interpreting,discussing}. We
recall that this quantity is affected by large uncertainties
\cite{uncert2} with $\left({g_{d,\rm max}}/{g_{d,\rm ref}}\right)^2 =
3.0$ and $\left({g_{d,\rm min}}/{g_{d,\rm ref}}\right)^2 = 0.12$
\cite{interpreting,discussing}.

We also calculate $\sigmavint$ which is the integral of $\sigmav$ from the present temperature
up to the freeze--out temperature $T_f$, since this quantity enters the neutralino relic abundance
(and, for dominant s--wave annihilation, implies $\sigmavint \equiv x_f \sigmav$):

\begin{equation}
\Omega_{\chi} h^2 = \frac{x_f}{{g_{\star}(x_f)}^{1/2}} \frac{9.9 \cdot
10^{-28} \; {\rm cm}^3 {\rm s}^{-1}}{\sigmavint},
\label{omega}
\end{equation}
where $x_f$ is defined as $x_f \equiv m_{\chi}/T_f$ and
${g_{\star}(x_f)}$ denotes the relativistic degrees of freedom of the
thermodynamic bath at $x_f$.

The values of $\Omega_{\chi} h^2$, as obtained from Eq. (\ref{omega}),
are employed to exclude neutralino configurations which would provide
values exceeding the upper bound for cold dark matter (CDM),
$(\Omega_{CDM} h^2)_{\rm max}$, and to rescale the local neutralino
density $\rho_{\chi}$, when $\Omega_{\chi} h^2$ turns out to be below
the lower bound for CDM, $(\Omega_{CDM} h^2)_{\rm min}$. In the latter
case, we rescale $\rho_{\chi}$ by the factor $\xi = \rho_{\chi}
/\rho_0$, where $\rho_0$ is the total local DM density ; $\xi$ is
conveniently taken as $\xi = {\rm min}\{1, \Omega_{\chi}
h^2/(\Omega_{CDM} h^2)_{\rm min}\}$ \cite{gaisser}.

In the present analysis, $(\Omega_{CDM} h^2)_{\rm min}$ and
$(\Omega_{CDM} h^2)_{\rm max}$ are assigned the values: $(\Omega_{CDM}
h^2)_{\rm min} = 0.11$, $(\Omega_{CDM} h^2)_{\rm max} = 0.13$ to
conform to the new measurements by the Planck Collaboration
\cite{planck}.

Since the rates of DM direct detection and those of processes due to pair annihilation (with the exclusion of processes taking place in macroscopic bodies) are proportional to
$\rho_{\chi}$ and $\rho_{\chi}^2$, respectively, in the following we will consider the quantities
 $\xi \sigma_{\rm scalar}^{(\rm nucleon)}$ and $\xi^2 \sigmav$.

 We calculate Higgs-boson masses and production cross sections using
 FeynHiggs \cite{feynhiggs}.

\section{Constraints}
\label{sec:constraints}

We give here a listing of requirements and constraints derived from a rich
set of experimental data.
In Sects. \ref{sec:lep} -- \ref{sec:colliders} are indicated the requirements
which are essential to qualify the model we are considering.
Sect. \ref{sec:bfact} reports other constraints
which can potentially bound the physical region of the supersymmetric
parameter space, but whose implications for our model are more involved and
thus possibly less compelling; we will also explicitly consider the possibility
of relaxing some of them. This aspect will be discussed later on.

\subsection{Constraints from the CERN $e^+ e^-$ collider LEP2}
\label{sec:lep}

These constraints take into account all data on supersymmetric and Higgs boson
searches \cite{LEPb} done at LEP2 (some of which are improved by those
obtained at the Tevatron and LHC, as discussed in the next subsection) as well as
 the upper bound on the invisible width for
the decay of the $Z$--boson into non Standard Model particles:
$\Gamma(Z \rightarrow \chi \chi) <$ 3 MeV \cite{aleph05,pdg}.

\subsection{Constraints from the Tevatron and  the LHC}
\label{sec:colliders}

Bounds on searches for supersymmetry from Tevatron and LHC are implemented as schematically
outlined below. The observation of a Higgs--like particle seen at the LHC imposes specific requirements
on the signal strengths factors for the production and decay of this boson, which have been applied
as discussed below.

%%%%%%%%%%
\smallskip
{\bf Signal strength factors for Higgs production/decay}.
 In the spirit of the present analysis, in the scanning of the supersymmetric parameter space
we select configurations which satisfy the following requirements, as established by the most
recent results at LHC \cite{moriond}:

\begin{eqnarray}
0.61 ~< ~R_{\gamma \gamma}~ <~ 1.57
\label{1}\\
0.75 ~< ~R_{Z Z}~ <~ 1.47
\label{2}\\
0.44 ~< ~R_{W W}~ <~ 1.24
\label{3}\\
0.21 ~< ~R_{\tau \tau}~ <~ 1.90,
\label{4}
\end{eqnarray}
where the ratio $R_{\gamma \gamma}$ is defined as:

\begin{equation}
R_{\gamma \gamma} = \frac{\sigma(p + p \rightarrow H_{125}) BR(H_{125} \rightarrow \gamma + \gamma)}
{\sigma_{SM}(p + p \rightarrow H_{125}) BR_{SM}(H_{125} \rightarrow \gamma + \gamma)},
\label{R1}
\end{equation}
and similarly for the other final states. Notice that the ranges of Eqs. (\ref{1}--\ref{4}) are $2 \sigma$
intervals.

%%%%%
\smallskip {\bf Bounds from search for Higgs decaying to tau
  pairs}. An upper bound in the plane $m_A - \tan \beta$ is obtained,
in our model, in an indirect way from the data reported by the CMS
Collaboration in Ref. \cite{CMS-12-050}.  A consistency check of our
procedure has been performed, by using the upper bound on the
production cross section reported in Ref. \cite{CMS-12-029} to obtain
the corresponding upper bound in the plane $m_A - \tan \beta$.
% in the
%specific case of the supersymmetric model employed there (which is
%different from typical LHC analyses).

%%%%%
\smallskip
{\bf Bounds on squark masses of the first two families and on the sbottom mass}. These bounds
are taken from the CMS official analysis of Ref. \cite{CMS-12-028}.

%%%%%
\smallskip
{\bf Bounds on the stop mass}. These bounds are taken from the official ATLAS analyses of Ref. \cite{ATLAS-1208.2590} for heavy stops and of
Ref. \cite{ATLAS-1208.4305} for light stops.

%%%%%
\smallskip
{\bf Decay $\bf B_s \bf \rightarrow \bf \mu^+ + \mu^-$}. We implement the constraint recently
derived by the LHCb Collaboration in Ref.  \cite{LHCb}:
\begin{equation}
1.1 \times 10^{-9} < BR(B_s \rightarrow \mu^+ + \mu^-) < 6.4 \times 10^{-9}
\label{BR}
\end{equation}
This is a 95 \% C.L. limit.

%%%%%
\smallskip
{\bf Search for the decay $\bf t \rightarrow b + H^+$}.

Whenever relevant, we have adopted the ATLAS 2--$\sigma$ upper bound on the branching ratio B($t
\rightarrow b + H^+$) as reported in Ref. \cite{ATLAS_tbh}.

%%%%%%%%%%
\subsection{Constraints from B factories and from (g - 2)$_{\mu}$ measurements}
\label{sec:bfact}

Flavor physics experiments are providing stringent bounds on many physical processes that
can be sizably affected by supersymmetric virtual corrections. Here we list the most relevant
ones for our analysis in the specific model we are assuming.

%%%%%
\smallskip
{\bf Measurement of the branching ratio of $b \rightarrow s + \gamma$}.
The rate for the branching ratio of the process $b \rightarrow s + \gamma$ is taken here as
 $2.89 \times 10^{-4} < BR(b \rightarrow s \gamma) < 4.21 \times 10^{-4}$.
This interval is larger by 25\% with
respect to the experimental determination \cite{bsgamma} in order to
take into account theoretical uncertainties in the supersymmetric
 contributions \cite{bsgamma_theorySUSY} to the branching ratio
of the process. For the Standard Model calculation, we employ the NNLO
results from Ref. \cite{bsgamma_theorySM}.

\smallskip
{\bf Search for the decay $B \rightarrow \tau + \nu$}.
We use here the range $0.38 \times 10^{-4} < BR(B \rightarrow \tau + \nu) < 1.42 \times 10^{-4}$
(world average at 95 \% C.L.) \cite{yy}.

\smallskip
{\bf Search for the decay $B \rightarrow D + \tau + \nu$}. A new range for the quantity
$R(D) \equiv BR(B \rightarrow D \tau \nu)/BR(B \rightarrow D e \nu)$
has been established by the BABAR Collaboration \cite{babar}:
$30.0 \times 10^{-2} < R(D) < 58.8 \times 10^{-2}$ (2$\sigma$ interval).

\smallskip {\bf Muon anomalous magnetic moment $(g - 2)_{\mu}$}. We
take the conservative 2$\sigma$ range $3.1 \times 10^{-10} \leq \Delta
a_{\mu} \leq 47.9 \times 10^{-10}$ for the deviation $\Delta a_{\mu}
\equiv a_{\mu}^{\rm exp} - a_{\mu}^{\rm the}$ of the experimental
world average of $a_\mu \equiv (g_{\mu} - 2)/2$\cite{bennet} from the
theoretical evaluation\cite{davier} (in the latter we estimate the
leading hadronic vacuum polarization contribution in the Standard
Model by combining the two determinations estimated from $e^{+}e^{-}$
and $\tau$-decay data).  We evaluate the supersymmetric contributions
to the muon anomalous magnetic moment within the MSSM by using the
formulae in Ref. \cite{moroi}.

\section{Selection of supersymmetric configurations and neutralino
  relic abundance}
\label{sec:results}

Here we provide the results of our analysis for scenarios I and II,
regarding the selection of supersymmetric configurations and the
neutralino relic abundance. The mass interval for the LHC Higgs--like
particle is taken here as 123 GeV $\leq m_{H_{125}} \leq$ 129 GeV.

\subsection{Scenario I: $H \equiv H_{125}$}
\label{sec:scenarioI}

%\begin{table}[t]
%\begin{center}
%\begin{tabular}{|c|c|}
%\hline
%\multicolumn{2}{|c|}{\bf Scenario I} \\
%\hline
%$\tan\beta$ 	& (4, 8) \\
%$\mu $  		& (1500, 2000) GeV \\
%$M_1$		& (5, 120) GeV \\
%$M_2$		& (100, 1000) GeV \\
%$M_3$		& $\sim$ 2000 GeV \\
%$m_{\tilde q_{12}}$	& (700, 2000) GeV \\
%$m_{\tilde t}$		& (700, 1000) GeV \\
%$m_{\tilde l_{12,L}}, m_{\tilde l_{12,R}}, m_{\tilde {\tau}_L}, m_{\tilde {\tau}_R}$  & (80, 200) GeV \\
%$m_A$		& (110, 150) GeV \\
%$|A|$			& (1.5, 3) \\
%\hline
%\end{tabular}
%\end{center}
%\caption{\label{tab:scenarioI}
%Values and intervals for the MSSM parameters outlined by the LHC bounds for Scenario I.}
%\end{table}

\begin{table}[t]
\begin{center}
\begin{tabular}{|c|c|}
\hline
\multicolumn{2}{|c|}{\bf Scenario I} \\
\hline
$\tan\beta$ 	& (4, 6) \\
$\mu $  		& (1800, 2000) GeV \\
$M_1$		& (40, 80) GeV \\
$M_2$		& (180, 800) GeV \\
$M_3$		& $\sim$ 2000 GeV \\
$m_{\tilde q_{12}}$	& (1400, 1600) GeV \\
$m_{\tilde t}$		& (1400, 1600) GeV \\
$m_{\tilde l_{12,L}}, m_{\tilde l_{12,R}}$  & $\sim$ 500 GeV \\
$m_{\tilde {\tau}_L}, m_{\tilde {\tau}_R}$  & (120, 200) GeV \\
$m_A$		& (100, 120) GeV \\
$|A|$			& (2.5, 2.8) \\
\hline
\end{tabular}
\end{center}
\caption{\label{tab:scenarioI}
Values and intervals for the MSSM parameters outlined by the LHC bounds for Scenario I.}
\end{table}

This scenario is defined by identifying the heavier
CP--even Higgs neutral boson $H$ of the MSSM with the LHC Higgs--like particle,
 {\it i.e.} $H \equiv H_{125}$. This implies that the mass interval for $H_{125}$
 obtained at the LHC has to be attained by $H$ (123 GeV $\leq m_H \leq$ 129 GeV),
 and this entails  $m_h \sim m_A \simeq (90 - 100)$ GeV,
 as already noticed above.
 The LHC constraints on the production rates in the various channels
 detailed in Eqs. (\ref{1}--\ref{4}) select the sector of
 supersymmetric parameter space reported in Table \ref{tab:scenarioI}.
 \footnote{For the lower bound on the slepton masses we use here the
   LEP values $m_{\tilde l} \gsim$ 80-100 GeV (depending on flavour)
   \cite{pdg}. These lower bounds actually depend on the condition
   that $m_{\tilde l} - m_{\chi_1} >$ O(3--15) GeV. If these
   conditions are not met, the slepton lower bound can decrease to
   about 40 GeV, with relevant implications for the neutralino
   phenomenology, as discussed in Ref. \cite{boehm}.}

The most peculiar feature of this region is represented by the high
values of the parameter $\mu$, a property which agrees with the
findings of Refs. \cite{hsw,b,carena4} and appears to be related to
the constraint imposed by Eq. (\ref{1}) on $R_{\gamma \gamma}$, as
remarked in Ref. \cite{b}. We also notice that the sector of parameter space defined in
Table \ref{tab:scenarioI} has some similarities with the scenario denoted by {\it
  low-$M_H$} in Ref. \cite{carena4}, though it differs in one
important feature: in our case the slepton masses (most notably the
mass parameters for $\tilde {\tau}_L$ and $\tilde {\tau}_R$) are
significantly lower. These are the prerequisites for having
configurations where the neutralino relic abundance does not exceed
the cosmological bound.

%%%%%%%%%%%%%%%%%%%%%%%%%%%%%%%%%%%%%%%%%%%%%%%%%%%%%%%%%%%%%%%
\begin{figure}[t]
\begin{center}
\includegraphics[width=1.00\linewidth, bb=55 200 526 634]{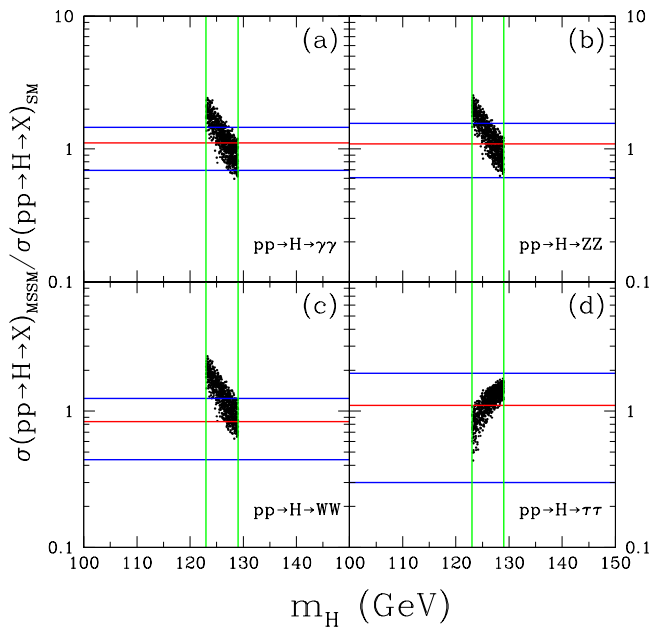}
\end{center}
\caption{Scenario I -- Signal strength factors as defined in
  Eq. (\protect\ref{R1}) for the production and decay of the heavy
  Higgs $H$ when 123 GeV$\le m_H\le$ 129 GeV). (a):
  $R_{\gamma\gamma}$; (b): $R_{ZZ}$; (c): $R_{WW}$; (d):
  $R_{\tau\tau}$.  The horizontal lines denote the allowed intervals
  obtained at the LHC, and given in Eq.(\protect\ref{1},\ref{2},\ref{3},\ref{4}).}
\label{HH_decays}
\end{figure}
%%%%%%%%%%%%%%%%%%%%%%%%%%%%%%%%%%%%%%%%%%%%%%%%%%%%%%%%%%%%%%%%

\begin{figure}[t]
\begin{center}
\includegraphics[width=1.00\linewidth, bb=55 200 526 634]{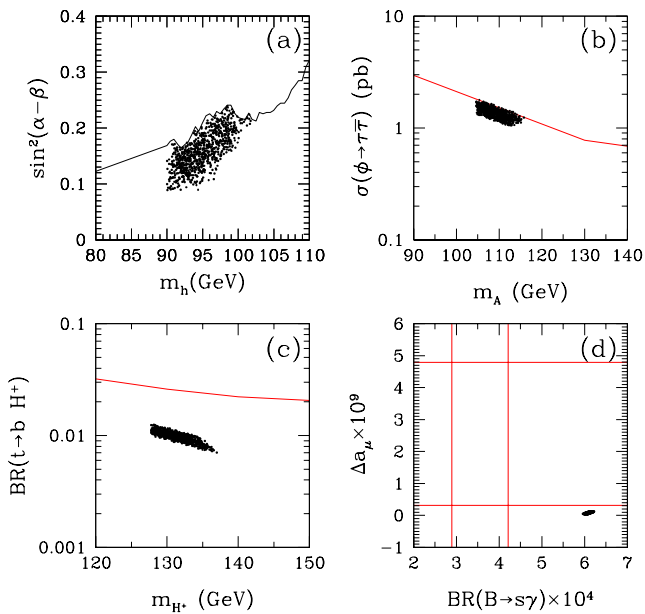}
\end{center}
\caption{Scenario I -- Some of the experimental constraints discussed
  in Section \protect\ref{sec:constraints} are compared to the
  corresponding theoretical expectations for the supersymmetric
  configurations reported in Table \ref{tab:scenarioI} and
  Fig. \protect\ref{HH_decays}. Panels (a)--(c) correspond to collider
  constraints: (a) LEP bound on the higgs production cross--section,
  reported in terms of the coupling factor $\sin^2(\alpha-\beta)$
  \protect\cite{LEPb}; (b) CMS bound on Higgs production and
  subsequent decay into $\tau\bar{\tau}$ \protect\cite{CMS-12-050}
  (the production cross section refers to $\phi=A$ unless $m_A\simeq
  m_h$ or $m_A\simeq m_H$, in which case $\phi=A,h$ or $\phi=A,H$,
  respectively); (c) ATLAS upper bound on the branching ratio
  BR$(t\rightarrow b H^{+})$ \protect\cite{ATLAS_tbh}. Panel (d) shows
  the extent of deviations from the two constraints on BR($b
  \rightarrow s + \gamma$) and $(g - 2)_{\mu}$.}
\label{HH_bounds}
\end{figure}

\begin{figure}[t]
\begin{center}
\includegraphics[width=1.00\linewidth, bb=55 200 526 634]{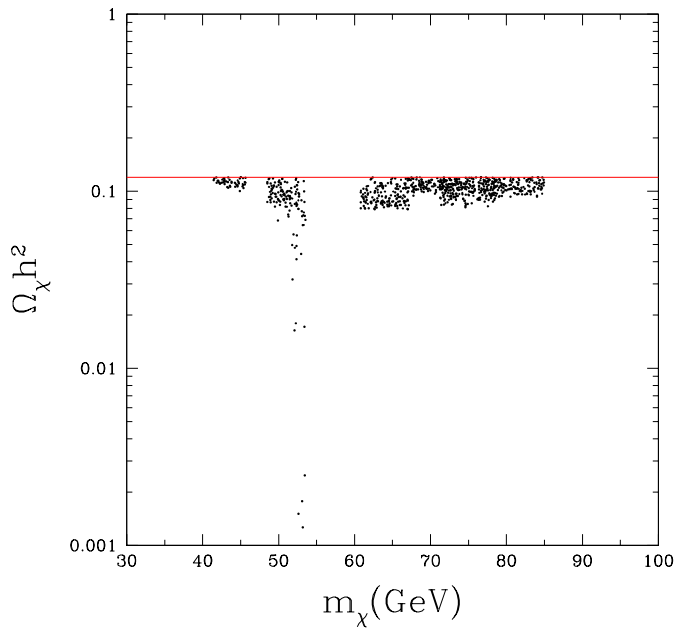}
\end{center}
\caption{Scenario I -- Neutralino relic abundance as a function of the
  neutralino mass for the supersymmetric configurations reported in
  Table \ref{tab:scenarioI} and Fig.\protect\ref{HH_decays}. The
  horizontal line represents the upper bound $(\Omega_{CDM} h^2)_{\rm
    max} = 0.13$ from the Planck Collaboration \cite{planck} on the
  cold dark matter content in the Universe.}
\label{HH_omega}
\end{figure}

\begin{figure}[t]
\begin{center}
\includegraphics[width=1.00\linewidth, bb=55 200 526 634]{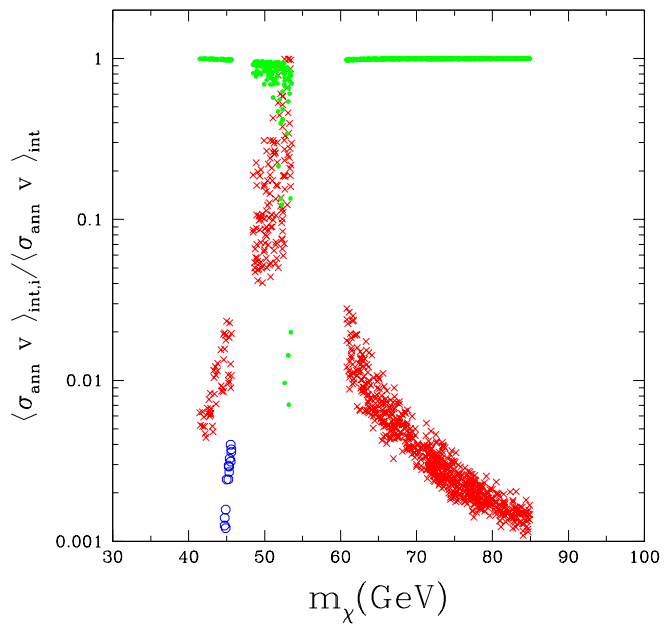}
\end{center}
\caption{Scenario I -- Fractional contributions of different
  annihilation channels to the integrated neutralino cross section
  times velocity $\sigmav_{int}$ for the supersymmetric configurations
  reported in Table \ref{tab:scenarioI} and
  Fig.\protect\ref{HH_decays}. (Green) dots: $\chi \chi\rightarrow
  f\bar{f}$ through slepton exchange; (red) crosses: $\chi
  \chi\rightarrow f\bar{f}$ through Higgs exchange; (blue) open
  circles: $\chi \chi\rightarrow f\bar{f}$ through $Z$--boson
  exchange.}
\label{HH_sigmav_contributions}
\end{figure}

The properties of the solutions we have found are displayed in
Figs. \ref{HH_decays} -- \ref{HH_sigmav_contributions}.  In
Fig. \ref{HH_decays} we show the signal strength factors for Higgs
production and decay at the LHC; we notice how a sizable subset of our
population of supersymmetric configurations fits quite well all LHC
data on these factors. This population satisfies also the other
relevant constraints from colliders as depicted in panels (a)--(c) of
Fig.\ref{HH_bounds}.  Panel (d) of Fig. \ref{HH_bounds}, instead,
shows that predictions for BR($b \rightarrow s + \gamma$) and $(g -
2)_{\mu}$ in scenario I deviate from the experimental bounds discussed
in Sect. \ref{sec:bfact}.  A minimal deviation occurs for $(g -
2)_{\mu}$, whereas a deviation of about 4$\sigma$ occurs for BR($b
\rightarrow s + \gamma$).  Therefore scenario I, which is perfectly
viable as far as accelerators data are concerned, is in tension with
the experimental bounds when also flavor physics determinations are
included (this will not be the case for scenario II, as discussed
below).  Contrary to accelerator physics constraints, these are
indirect bounds and rely to some degree of cancellation of various
terms \cite{discussing}, which may not be fully under theoretical
control. We therefore discuss the implications of scenario I for dark
matter, nevertheless reminding that this scenario exhibit a
significant level of tension with indirect bounds on supersymmetry.
%%%%%%%%%%%%%%%%%%%%%%%%%%%%%%%%%%%%%%%%%%%%%%%%%%%%%%%%%%%%%%%
%%%%%%%%%%%%%%%%%%%%%%%%%%%%%%%%%%%%%%%%%%%%%%%%%%%%%%%%%%%%%%%%

\begin{figure}[t]
\begin{center}
\includegraphics[width=1.00\linewidth, bb=55 200 526 634]{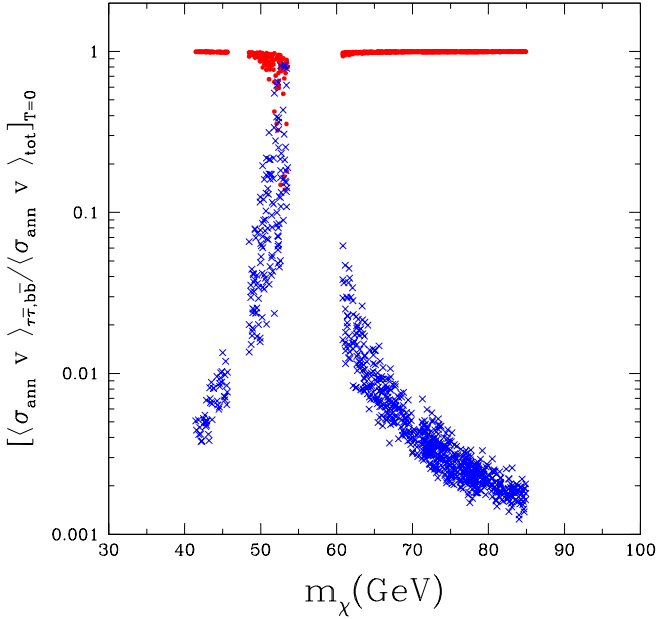}
\end{center}
\caption{Scenario I -- Ratios $[\sigmav_i/\sigmav_{tot}]_{T=0}$
  between the neutralino annihilation cross sections times velocity to
  the final states $i=\tau\tau,b\bar{b}$ and the total annihilation
  cross section times velocity, both calculated at zero temperature,
  for the supersymmetric configurations reported in Table
  \ref{tab:scenarioI} and Fig.\protect\ref{HH_decays}. (Red) dots:
  $\tau\bar{\tau}$ final state; (blue) crosses: $b \bar{b}$ final
  state.}
\label{HH_br_f_fbar}
\end{figure}

In the present scenario the neutralino mass sits in the range:
$m_{\chi} \simeq (40-85)$ GeV.  As shown in Fig. \ref{HH_omega}, most
of our configurations have a sizable neutralino relic
abundance. Fig. \ref{HH_sigmav_contributions} illustrates the
contributions of different annihilation channels to the integrated
cross section $\sigmav_{int}$; we notice that, as anticipated, light
sleptons are instrumental in keeping the annihilation cross section
large enough to comply with the experimental upper bound on
$\Omega_{\chi} h^2$, since diagrams with exchange of a slepton
dominate $\sigmav_{int}$ over the whole range of the available
neutralino masses, with the exception of a small mass range around
$m_{\chi}\simeq m_A/2$, where resonant annihilation through $A$
exchange can become important.  On the other hand, $Z$--boson exchange
remains sub--dominant even close to the pole in the corresponding
annihilation cross section, $m_{\chi}\simeq M_Z/2$, since the $Z$
boson couples to the neutralino only through its Higgsino components,
while in this scenario the neutralino is a Bino of extremely high
purity, due to the very large values required for the $\mu$ parameter,
as specified in Table \ref{tab:scenarioI}. Finally, for the same set
of supersymmetric configurations we show in Fig. \ref{HH_br_f_fbar}
the ratios $[\sigmav_i/\sigmav_{tot}]_{T=0}$ between the neutralino
annihilation cross sections times velocity to the final states
$i=\tau\tau,b\bar{b}$ and the total annihilation cross section times
velocity, both calculated at zero temperature. The latter quantities
are relevant for the evaluation of indirect signals, as discussed in
Section \ref{sec:indirect}. As shown in the plot, annihilation to
$\tau\bar{\tau}$ (driven by the exchange of light staus) is dominant,
with a sub-dominant contribution from the $b\bar{b}$ annihilation
channel (which, as in the case of $\sigmav_{int}$, can become sizeable
through resonant annihilation through $A$ exchange).
%%%%%%%%%%%%%%%%%%%%%%%%%%%%%%%%%%%%%%%%%%%%%%%%%%%%%%%%%%%%%%%
%%%%%%%%%%%%%%%%%%%%%%%%%%%%%%%%%%%%%%%%%%%%%%%%%%%%%%%%%%%%%%%%

%%%%%%%%%%%%%%%%%%%%%%%%%%%%%%%%%%%%%%%%%%%%%%%%%%%%%%%%%%%%%%%
%%%%%%%%%%%%%%%%%%%%%%%%%%%%%%%%%%%%%%%%%%%%%%%%%%%%%%%%%%%%%%%%

%%%%%%%%%%%%%%%%%%%%%%%%%%%%%%%%%%%%%%%%%%%%%%%%%%%%%%%%%%%%%%%
%%%%%%%%%%%%%%%%%%%%%%%%%%%%%%%%%%%%%%%%%%%%%%%%%%%%%%%%%%%%%%%%

\subsection{Scenario II: $h \equiv H_{125}$}
\label{sec:scenarioII}

%\begin{table}[t]
%\begin{center}
%\begin{tabular}{|c|c|}
%\hline
%\multicolumn{2}{|c|}{\bf Scenario II} \\
%\hline
%$\tan\beta$ 	& (4, 15) \\
%$|\mu|$  		& (100, 1000) GeV \\
%$M_1$		& (5, 120) GeV \\
%$M_2$		& (100, 1000) GeV \\
%$M_3$		& $\sim$ 2000 GeV \\
%$m_{\tilde q_{12}}$	& (700, 2000) GeV \\
%$m_{\tilde t}$		& (700, 1000) GeV \\
%$m_{\tilde l_{12,L}}, m_{\tilde l_{12,R}}, m_{\tilde {\tau}_L}, m_{\tilde {\tau}_R}$  & (80, 200) GeV \\
%$m_A$		& (200, 1000) GeV \\
%$|A|$			& (1.5, 3) \\
%\hline
%\end{tabular}
%\end{center}
%\caption{\label{tab:scenarioII}
%Values and intervals for the MSSM parameters outlined by the LHC bounds for Scenario II.}
%\end{table}

\begin{table}[t]
\begin{center}
\begin{tabular}{|c|c|}
\hline
\multicolumn{2}{|c|}{\bf Scenario II} \\
\hline
$\tan\beta$ 	& (4, 20) \\
$|\mu|$  		& (100, 400) GeV \\
$M_1$		& (40, 170) GeV \\
$M_2$		& (100, 1000) GeV \\
$M_3$		& $\sim$ 2000 GeV \\
$m_{\tilde q_{12}}$	& (700, 2000) GeV \\
$m_{\tilde t}$		& (700, 1200) GeV \\
$m_{\tilde l_{12,L}}, m_{\tilde l_{12,R}}, m_{\tilde {\tau}_L}, m_{\tilde {\tau}_R}$  & (80, 1000) GeV \\
$m_A$		& (200, 1000) GeV \\
$|A|$			& (1.5, 2.5) \\
\hline
\end{tabular}
\end{center}
\caption{\label{tab:scenarioII}
Values and intervals for the MSSM parameters outlined by the LHC bounds for Scenario II.}
\end{table}

\begin{figure}[t]
\begin{center}
\includegraphics[width=1.00\linewidth, bb=50 48 525 506]{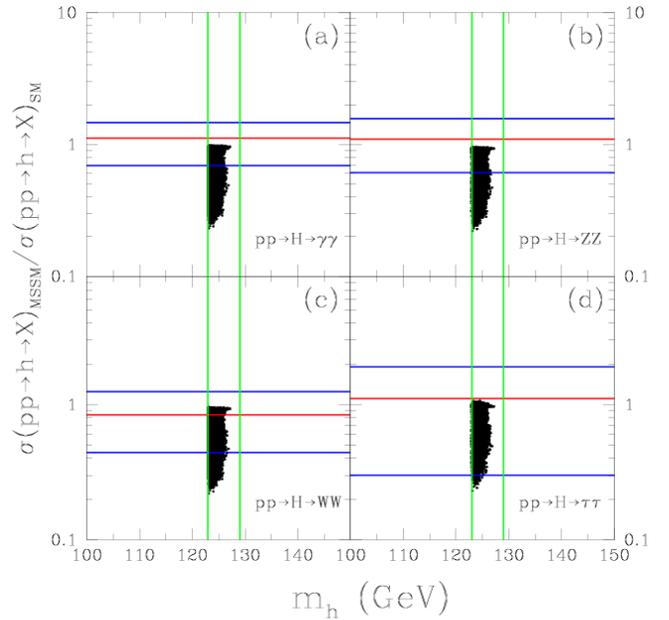}
\end{center}
\caption{Scenario II - The same as in Fig. \protect\ref{HH_decays}, except that here 123 GeV$\le
  m_h\le$ 129 GeV.}
\label{h0_decays}
\end{figure}

\begin{figure}[t]
\begin{center}
\includegraphics[width=1.0\linewidth, bb=50 50 504 508]{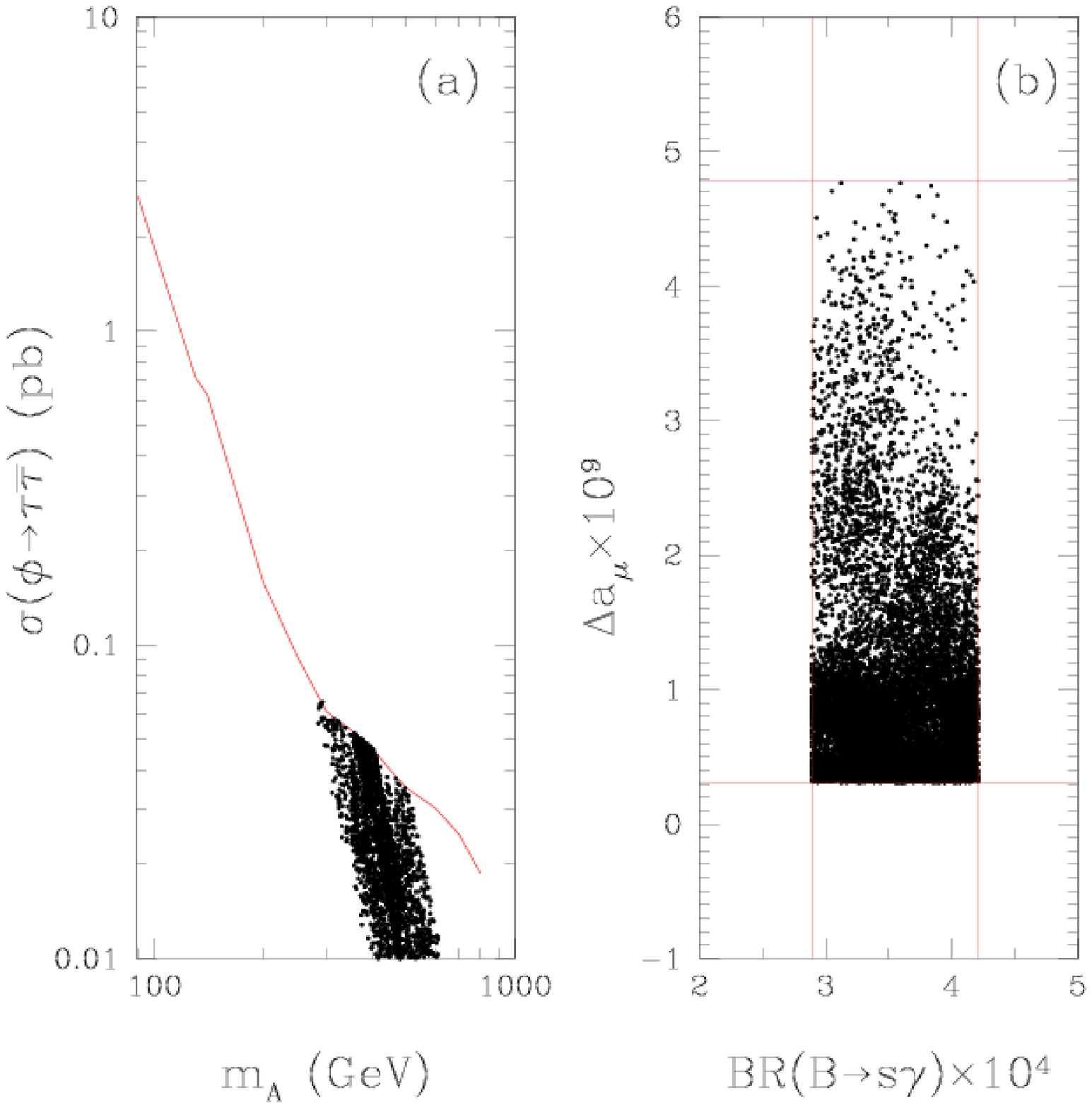}
\end{center}
\caption{Scenario II - Two of the experimental constraints discussed
  in Section \protect\ref{sec:constraints} are compared to the
  corresponding theoretical expectations for the supersymmtric
  configurations reported in Table \ref{tab:scenarioII} and in
  Fig. \protect\ref{h0_decays}. (a) CMS bound on Higgs production and
  subsequent decay into $\tau\bar{\tau}$ \protect\cite{CMS-12-050}
  (the production cross section refers to $\phi=A$ unless $m_A\simeq
  m_h$ or $m_A\simeq m_H$, in which case $\phi=A,h$ or $\phi=A,H$,
  respectively); (b) the two constraints on BR($b \rightarrow s +
  \gamma$) and $(g - 2)_{\mu}$, which in this scenario are
  simultaneously satisfied.}
\label{h0_bounds}
\end{figure}

\begin{figure}[t]
\begin{center}
\includegraphics[width=1.00\linewidth, bb=20 48 502 506]{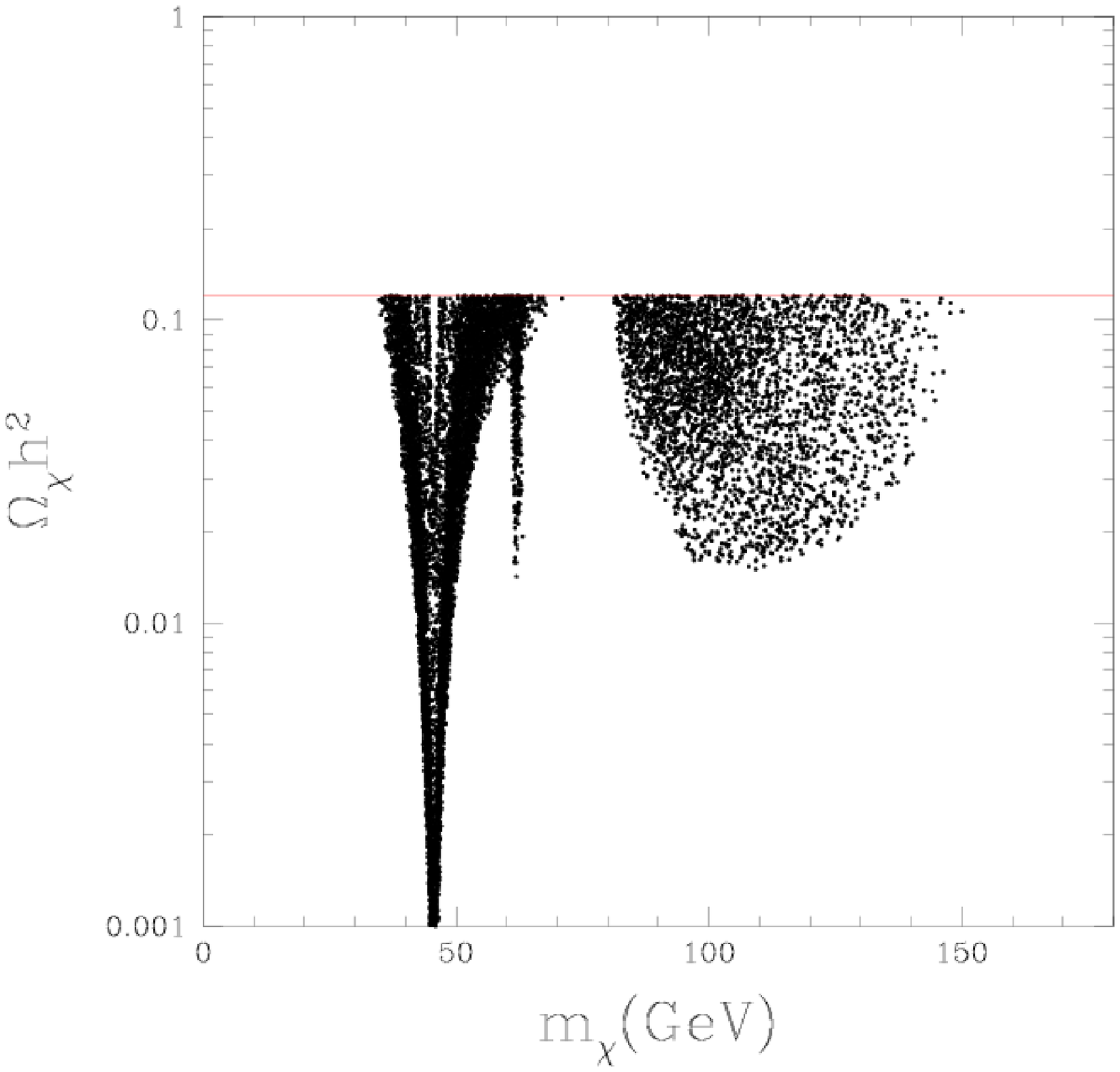}
\end{center}
\caption{Scenario II - The same as in Fig.\protect\ref{HH_omega} for the
  supersymmetric configurations reported in Table \ref{tab:scenarioII} and in  Fig. \protect\ref{h0_decays} (Scenario
  II in the text).}
\label{h0_omega}
\end{figure}

\begin{figure}[t]
\begin{center}
\includegraphics[width=1.00\linewidth, bb=30 48 512 504]{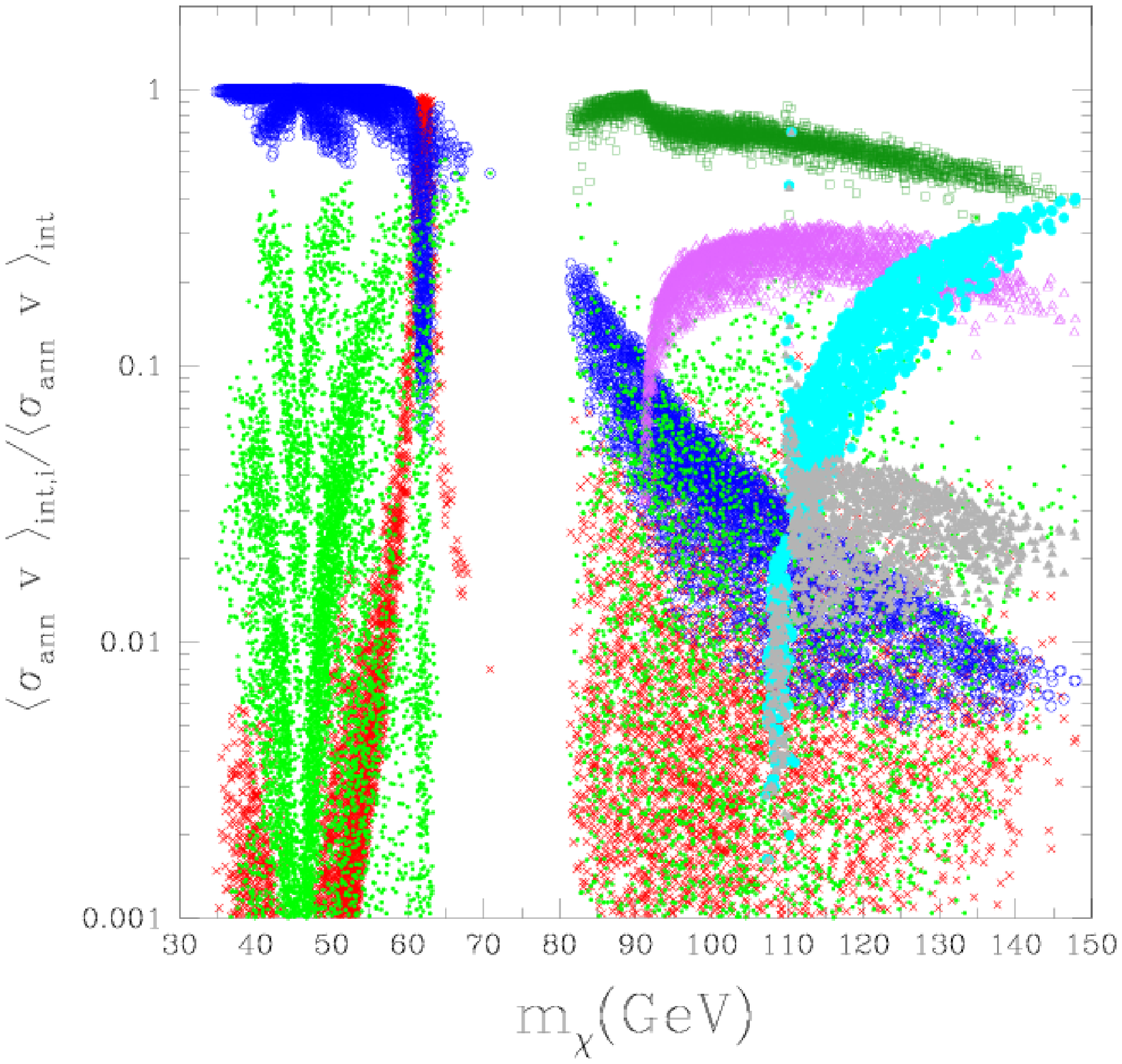}
\end{center}
\caption{Scenario II - The same as in
  Fig. \protect\ref{HH_sigmav_contributions}, for the supersymmetric
  configurations reported in Table \ref{tab:scenarioII} and in
  Fig. \protect\ref{h0_decays}.  (Green) dots: $\chi \chi\rightarrow
  f\bar{f}$ through slepton exchange; (red) crosses: $\chi
  \chi\rightarrow f\bar{f}$ through Higgs exchange; (blue) open
  circles: $\chi \chi\rightarrow f\bar{f}$ through $Z$--boson
  exchange; (dark green) open squares: $\chi \chi\rightarrow WW$;
  (purple) open triangles: $\chi \chi\rightarrow ZZ$; (cyan) filled
  circles: $\chi \chi\rightarrow Zh$; (grey) filled triangles: $\chi
  \chi\rightarrow hh$.}
\label{h0_sigmav_contributions}
\end{figure}

\begin{figure}[t]
\begin{center}
\includegraphics[width=1.00\linewidth, bb=20 48 512 504]{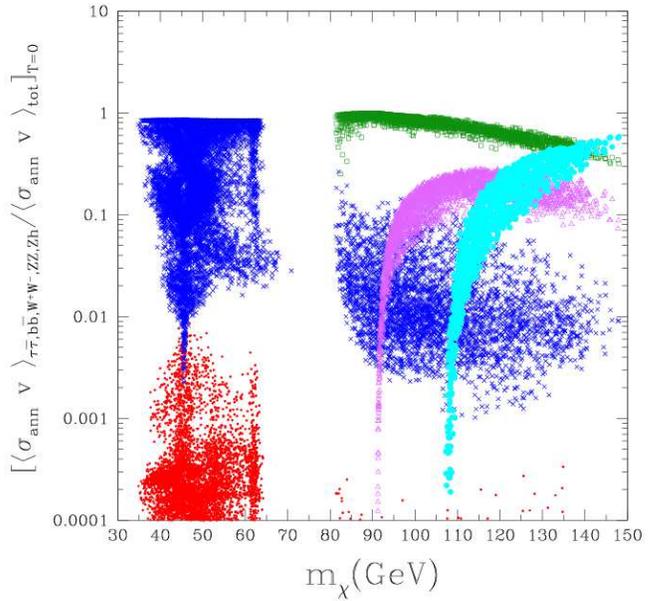}
\end{center}
\caption{Scenario II - The same as in Fig. \protect\ref{HH_br_f_fbar},
  for the supersymmetric configurations reported in Table
  \ref{tab:scenarioII} and in Fig. \protect\ref{h0_decays}. (Red)
  dots: $\tau\bar{\tau}$ final state; (blue) crosses: $b\bar{b}$ final
  state; (dark green) open squares: $W^{+}W^{-}$ final state; (purple)
  open triangles: $ZZ$ final state; (cyan) filled circles: $Zh$ final
  state.}
\label{h0_br_f_fbar}
\end{figure}

This scenario is defined by the alternative choice, that identifies
the lighter CP--even Higgs neutral boson $h$ of the MSSM with the LHC
Higgs--like particle, {\it i.e.} $h \equiv H_{125}$. This therefore
implies: 123 GeV $\leq m_h \leq$ 129 GeV.
The scan of the MSSM parameter space that produces a population of
configurations satisfying all requirements and constraints mentioned
in Sect. \ref{sec:constraints} identifies the sector outlined in Table
\ref{tab:scenarioII}.  The features of this population are displayed
in Figs. \ref{h0_decays} -- \ref{h0_br_f_fbar}.

Fig. \ref{h0_decays} shows how the
requirements for the signal strength factors are verified for our
configurations.  The constraint derived from LHC searches for Higgs
decaying to a tau pair implies for the mass of the CP--odd Higgs $A$ the
lower bound: $m_A \gsim$ 300 GeV, as indicated by panel (a) of Fig. \ref{h0_bounds}.
Fig. \ref{h0_bounds}(b) shows that, at variance with the previous case, in 
scenario II the constraints BR($b \rightarrow s + \gamma$) and $(g -
2)_{\mu}$ are satisfied. It is worth noting that also the bound on the
branching ratio for the invisible decay $h \rightarrow \chi + \chi$
\cite{inv}, not explicitly discussed before, is respected.

%%%%%%%%%%%%%%%%%%%%%%%%%%%%%%%%%%%%%%%%%%%%%%%%%%%%%%%%%%%%%%%
%%%%%%%%%%%%%%%%%%%%%%%%%%%%%%%%%%%%%%%%%%%%%%%%%%%%%%%%%%%%%%%%

%%%%%%%%%%%%%%%%%%%%%%%%%%%%%%%%%%%%%%%%%%%%%%%%%%%%%%%%%%%%%%%
%%%%%%%%%%%%%%%%%%%%%%%%%%%%%%%%%%%%%%%%%%%%%%%%%%%%%%%%%%%%%%%%

The plot of Fig. \ref{h0_omega}, displaying the neutralino relic
abundance versus the neutralino mass, shows that $m_{\chi}$ has the
lower limit $m_{\chi} \gsim$ 30 GeV and that there exists a break in
the range 70 GeV $\lsim m_{\chi} \lsim$ 85 GeV, this interval being
disallowed by the requirement that $\Omega_{\chi} h^2 \leq
(\Omega_{CDM} h^2)_{\rm max}$. In turn, this property is due to the
strong enhancement in the pair annihilation amplitude when $m_{\chi}$
runs over the values $\frac{1}{2} m_h, \frac{1}{2} m_Z$.

%%%%%%%%%%%%%%%%%%%%%%%%%%%%%%%%%%%%%%%%%%%%%%%%%%%%%%%%%%%%%%%
%%%%%%%%%%%%%%%%%%%%%%%%%%%%%%%%%%%%%%%%%%%%%%%%%%%%%%%%%%%%%%%%

Fig. \ref{h0_sigmav_contributions} displays the various contributions
to $\sigmav_{int}$ and shows that dominances in the annihilation
amplitude are as follows: a) dominance of annihilation to fermions
through $Z$--exchange in the range 30 GeV $\lsim m_{\chi} \lsim$ 60
GeV; b) dominance of annihilation to fermions through light scalar
Higgs--exchange for $m_{\chi} \simeq m_h/2$ ; c) dominance of
annihilation to $W^{+}W^{-}$ for $m_{\chi} > m_W$. To conclude the
discussion, Fig.\ref{h0_br_f_fbar} shows the ratios
$[\sigmav_i/\sigmav_{tot}]_{T=0}$ between the neutralino annihilation
cross section times velocity to the final states
$i=\tau\tau,b\bar{b},W^{+}W^{-},ZZ,Zh$ and the total annihilation
cross section times velocity, both calculated at zero temperature: as
shown in the plot, $\tau\bar{\tau}$ dominates when $m_{\chi}\lsim$ 65
GeV, $W^{+}W^{-}$ prevails when $m_W<m_{\chi}<m_Z+m_h$, and $Zh$
dominates at larger masses.

%%%%%%%%%%%%%%%%%%%%%%%%%%%%%%%%%%%%%%%%%%%%%%%%%%%%%%%%%%%%%%%
%%%%%%%%%%%%%%%%%%%%%%%%%%%%%%%%%%%%%%%%%%%%%%%%%%%%%%%%%%%%%%%%

%%%%%%%%%%%%%%%%%%%%%%%%%%%%%%%%%%%%%%%%%%%%%%%%%%%%%%%%%%%%%%%
%%%%%%%%%%%%%%%%%%%%%%%%%%%%%%%%%%%%%%%%%%%%%%%%%%%%%%%%%%%%%%%%

%%%%%%%%%%%%%%%%%%%%%%%%%%%%%%%%%%%%%%%%%%%%%%%%%%%%%%%%%%%%%%%
\begin{figure}[t]
\begin{center}
\includegraphics[width=1.00\linewidth, bb=28 48 510 505]{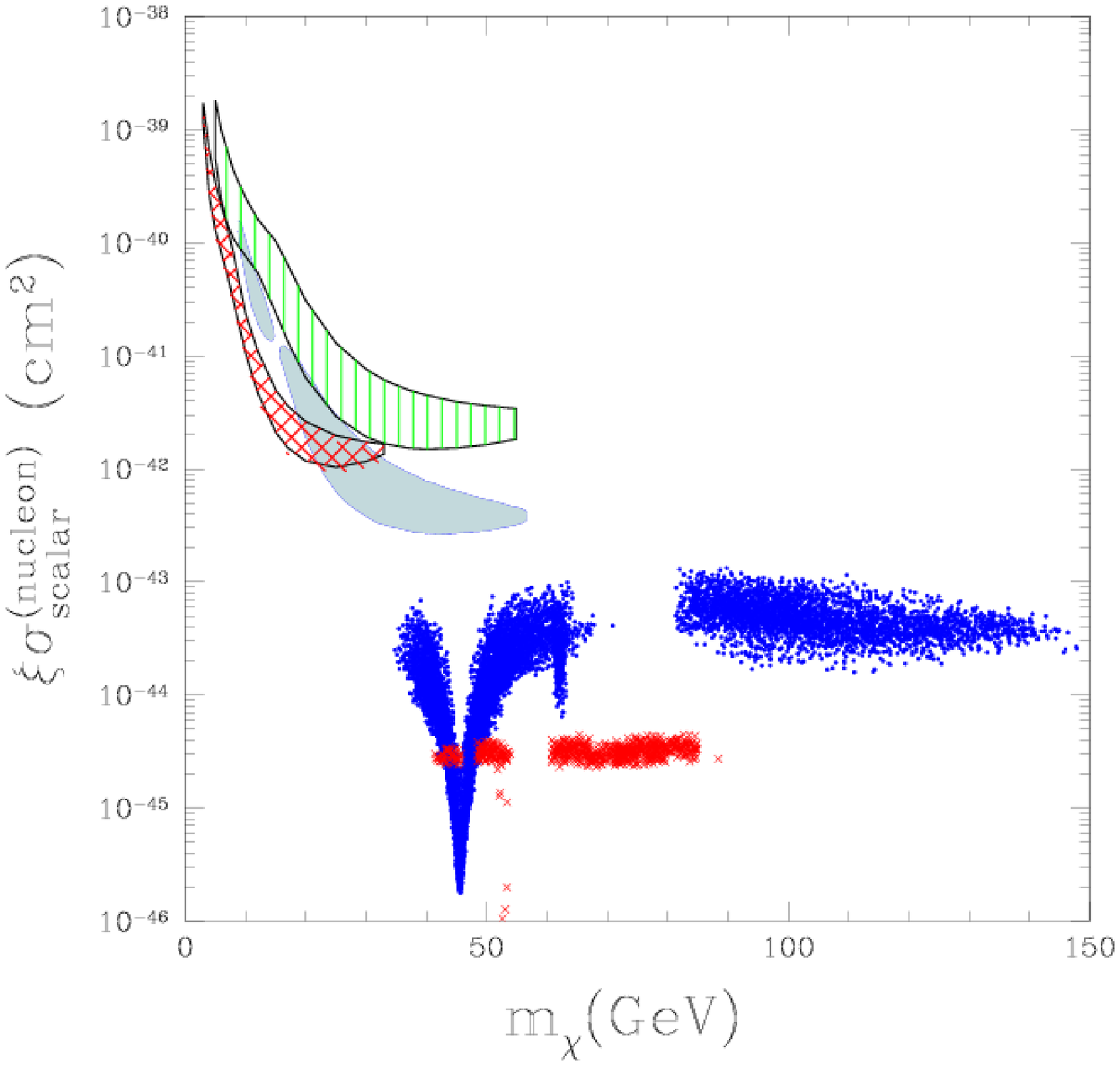}
\end{center}
\caption{Neutralino--nucleon coherent cross section times the
  rescaling factor $\xi \sigma_{\rm scalar}^{(\rm nucleon)}$. (Red)
  crosses: supersymmetric configurations plotted in
  Fig.\protect\ref{HH_decays} (Scenario I in the text); (blue) dots:
  supersymmetric configurations plotted in Fig.\protect\ref{h0_decays}
  (Scenario II in the text).  The hatched areas denote the DAMA/LIBRA
  annual modulation regions \protect\cite{dama2010}: the (green) vertically--hatched region
  refers to the case where constant values of 0.3 and 0.09 are taken
  for the quenching factors of Na and I, respectively\protect\cite{observ};
  the (red) crossed-hatched is obtained by using the
  energy--dependent Na and I quenching factors as established by the
  procedure given in Ref. \protect\cite{tretyak}. The gray regions are those
  compatible with the CRESST excess \protect\cite{cresst}. In all cases a
  possible channeling effect is not included.The halo distribution
  function used to extract the experimental regions is given in the
  text. For other distribution functions see \protect\cite{observ}}
  \label{HH_h0_sigma_direct}
\end{figure}
%%%%%%%%%%%%%%%%%%%%%%%%%%%%%%%%%%%%%%%%%%%%%%%%%%%%%%%%%%%%%%%%

\section{Direct detection.}
\label{sec:direct}

We turn now to the evaluation of the relevant quantity for DM direct
detection, $\xi \sigma_{\rm scalar}^{(\rm nucleon)}$. The values for
this quantity are shown in the scatter plot of
Fig. \ref{HH_h0_sigma_direct} together with the regions pertaining to
the signals measured by the experiments of DM direct detection of
Refs. \cite{dama2010,cresst} (other experimental results showing an
excess of events compatible with a positive signal are reported in
Refs. \cite{cogent,cdms2013}). In particular, in
Fig. \ref{HH_h0_sigma_direct} (red) crosses represent configurations
found in the set of Scenario I, while (blue) dots correspond to
configurations found in Scenario II.  The experimental domains shown
here were obtained by using for the velocity distribution function of
the galactic dark matter those pertaining to a standard isothermal
sphere with $\rho_0 = 0.30$ GeV cm$^{-3}$, $v_0 = 220$ km sec$^{-1}$,
with $v_{\rm esc} = 650$ km sec$^{-1}$ for the DAMA/LIBRA experiment
and $v_{\rm esc} = 544$ km sec$^{-1}$ for CRESST and for specific sets
of experimental parameters (quenching factors and others), as
discussed in Refs. \cite{dama2010,cresst}. Including
uncertainties of various origin, the experimental regions would expand
as indicated for instance in Fig. 7 of Ref.  \cite{observ}.

One notices that the set of configurations found in the scenario I
generate very low rates for direct detection of relic
neutralinos. Thus in this scheme neutralinos does not appear be
responsible for the signals measured by the experiments of DM direct
detection of Refs. \cite{dama2010,cresst,cogent,cdms2013}.

It is worth stressing that these conclusions rest heavily on the
results recently obtained from colliders; in particular, very
constraining are the conditions expressed in
Eqs. (\ref{1},\ref{2},\ref{3},\ref{4}) and the bounds implied by the
search for Higgs decay into tau pairs, that constrain the parameter
$\mu$ to be very large and $\tan \beta$ small. Should these
constraints significantly relax in the future, as a consequence of
further experimental data and analyses from colliders, the theoretical
values of $\xi \sigma_{\rm scalar}^{(\rm nucleon)}$ would compare to
the data of DM direct detection much more favorably, as for instance
depicted in Fig. 5 of Ref. \cite{phenomenology}.

In the case of scenario II, in view of the experimental uncertainties
mentioned above and of the theoretical uncertainties related to the
parameter $g_d$ (see Sect. \ref{sec:model}) , the gap between the
experimental regions and the scatter plot shown in Fig.
\ref{HH_h0_sigma_direct} could somewhat narrow down. Most of the
theoretical values shown in Fig.  \ref{HH_h0_sigma_direct} are in tension
with the experimental bounds given by other DM experiments (see for
instance Ref. \cite{xenon100,cdms}).

\section{Indirect detection.}
\label{sec:indirect}

In order to study the capability of indirect signals to probe neutralino dark matter in scenario I
and scenario II, we discuss the exotic component in cosmic rays represented by antiprotons, and
the contribution to the so--called isotropic gamma--ray background (IGRB) due to the production of 
gamma--rays at high latitudes from annihilation in our Galaxy.

\begin{figure*}[t]
\begin{center}
\includegraphics[width=0.49\linewidth]{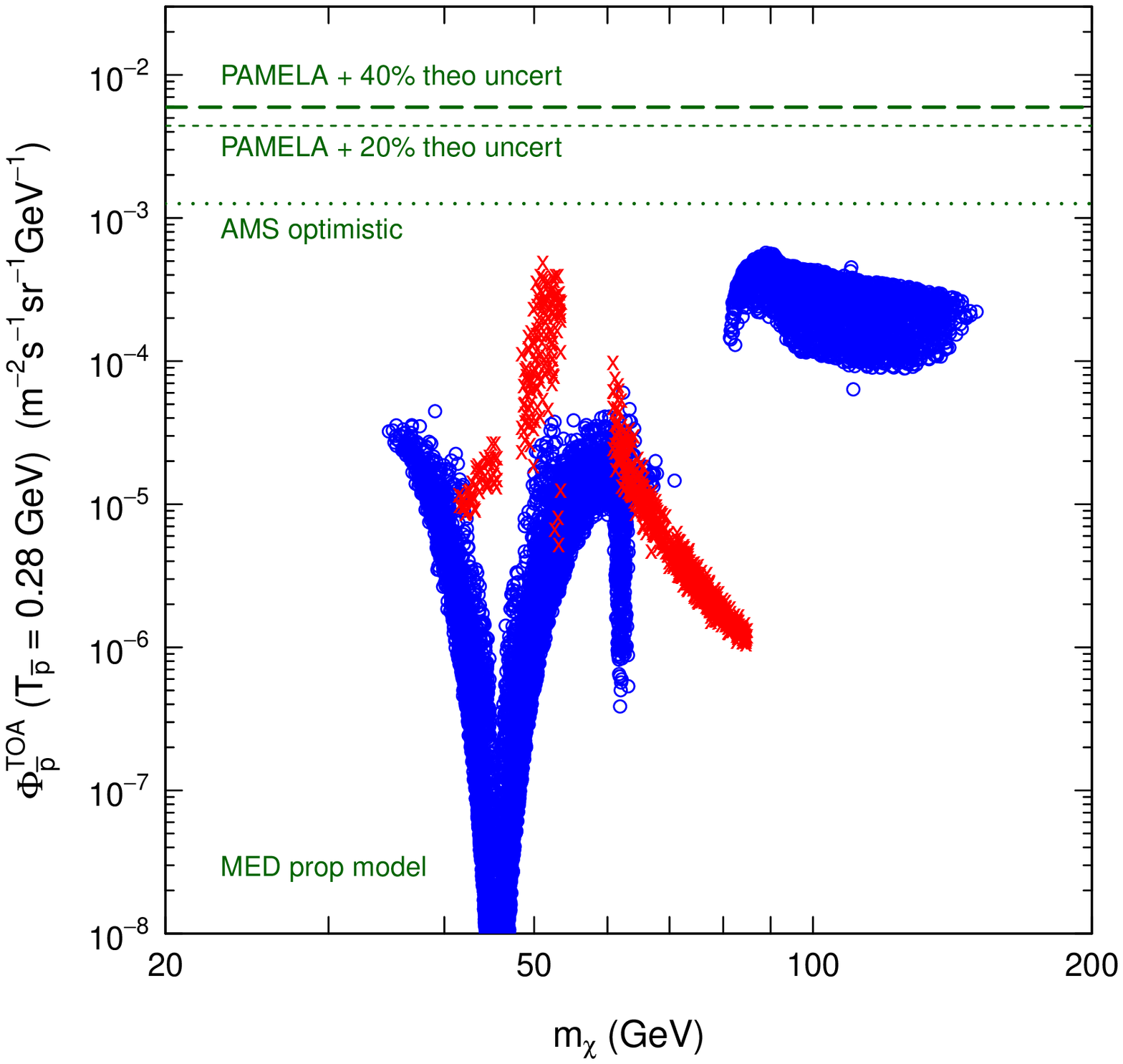}
\includegraphics[width=0.49\linewidth]{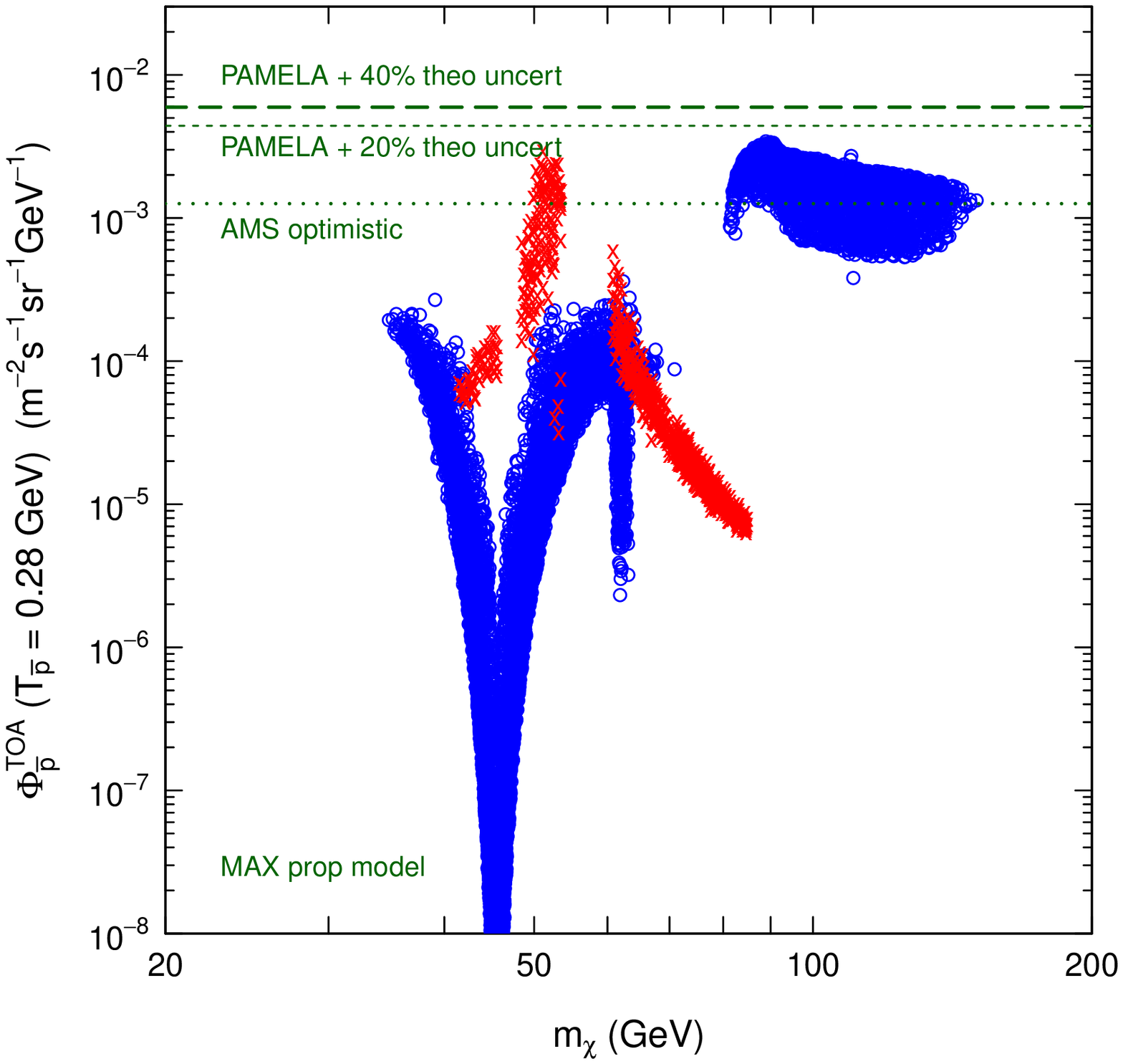}
\end{center}
\caption{Antiproton fluxes in the first PAMELA energy bin 
($T_{\bar p} = 0.28$ GeV) for the configurations of scenario I (red crosses) and scenario II (blue circles).
The upper long--dashed line denotes the 95\% C.L. bound by using the PAMELA \cite{Adriani:2010rc}
data and adding in quadrature a 40\% theoretical error on the theoretical determination of the antiproton background. The short--dashed line shows the same upper bound, with a 20\% estimate of the theoretical uncertainty.
The dotted line denotes the reaching capabilities of AMS \cite{AMS}, provided the total experimental and theoretical uncertainties are reduced to 10\%. The left panel refers to a galactic propagation model with the
MED values of propagation parameters \cite{Donato:2003xg}; the right panel refers to the MAX set of parameters \cite{Donato:2003xg}.
}
\label{fig:pbar028}
\end{figure*}

Antiprotons are potentially able to provide quite strong bounds on dark matter annihilation
in our Galaxy. We therefore calculate the antiproton production in both scenarios and compare
them with the PAMELA measurements of the absolute antiproton flux \cite{Adriani:2010rc}.
Similar bounds can be obtained with the BESS--Polar II determination \cite{Abe:2011nx}.
Fig. \ref{fig:pbar028} shows the antiproton fluxes in the first PAMELA energy bin 
($T_{\bar p} = 0.28$ GeV) for the configurations of scenario I (red crosses) and scenario II (blue circles).
The left panel refers to a galactic propagation model with the
MED values of propagation parameters \cite{Donato:2003xg}; the right panel refers to the MAX set of parameters \cite{Donato:2003xg}. The MAX set refers to the configuration in
the space of propagation parameters which provides the largest antiproton fluxes (mostly due to a
large volume of the cosmic--rays confinement region), while being
allowed by B/C measurements \cite{Donato:2003xg}.

The upper long--dashed line denotes the 95\% C.L. bound by using the
PAMELA data \cite{Adriani:2010rc} and adding in quadrature a 40\%
theoretical error on the theoretical determination of the antiproton
background. This generous allowance is taken under consideration
because of uncertainties in the nuclear cross sections relevant for
the secondary production \cite{Donato:2003xg}. The modification of the
bound when a smaller estimate of the theoretical uncertainty (20\%) is
adopted \cite{Donato:2003xg} is shown by the short--dashed line. We
notice that antiprotons are far from bounding the configurations of
both scenario I and II. This is due to the fact that the dominant
channel of annihilation in a large portion of the parameter space of
both scenarios is a leptonic one (namely, $\tau\bar\tau$) which is
unable to produce a relevant amount of antiprotons. Only those
configurations of scenario I where the $b\bar b$ final state dominates
(very few configurations with a neutralino mass close to 55 GeV, as
seen in Fig. \ref{HH_br_f_fbar}) and the configurations of scenario II
where the gauge--bosons final state (for neutralino masses above 80
GeV, as seen in Fig. \ref{h0_br_f_fbar}) dominates are able to produce
an antiproton flux that reaches its maximal values. Dominant hadronic
($b\bar b$) final states for neutralino masses below 70 GeV are
accompanied by small values of the neutralino relic abundance: this
has a strong impact in reducing the antiproton flux, due to the
squared appearance of the rescaling factor $\xi$ in indirect signals
(since they depend on $\rho_\chi^2$).

\begin{figure*}[t]
\begin{center}
\includegraphics[width=0.49\linewidth]{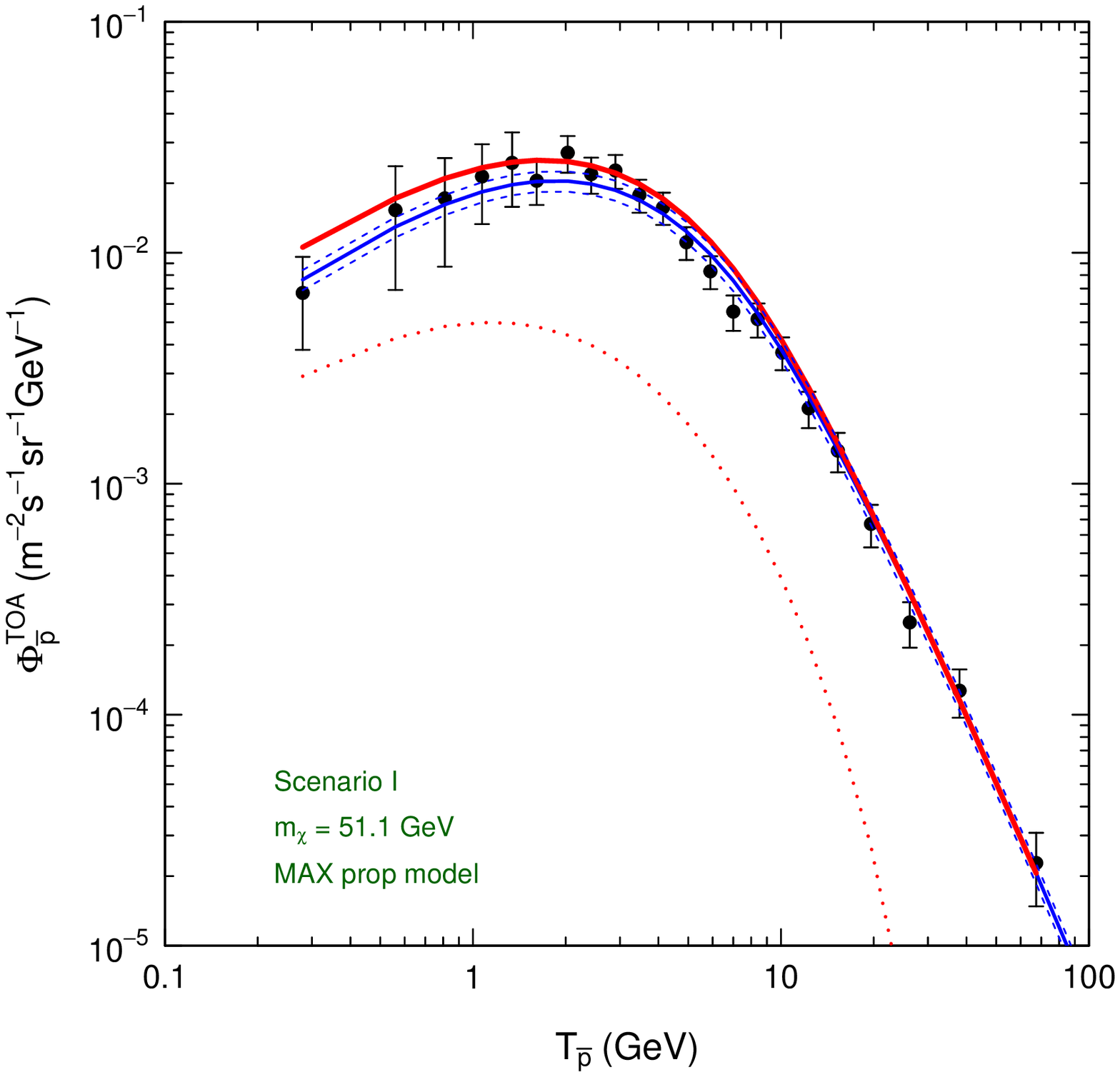}
\includegraphics[width=0.49\linewidth]{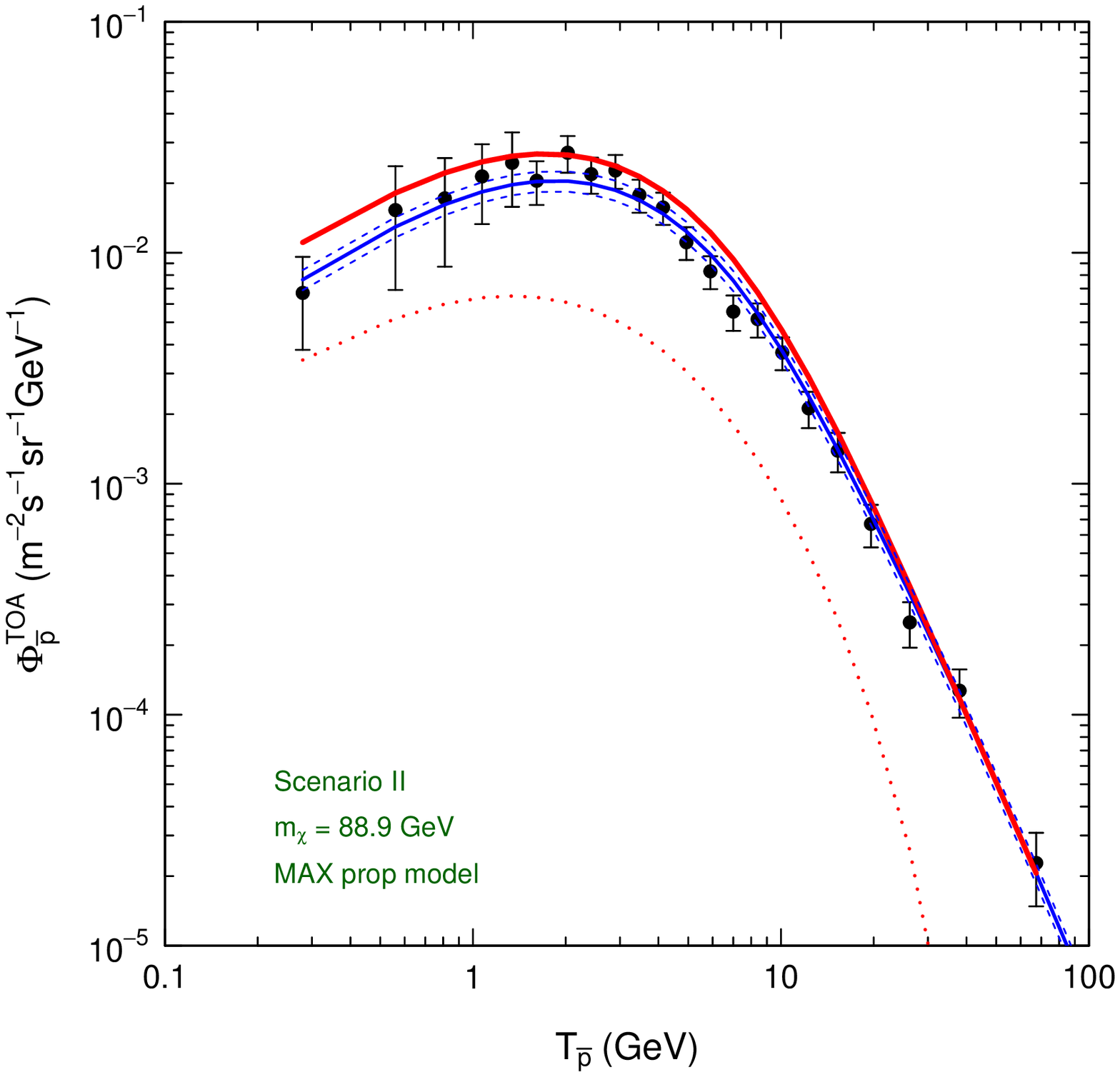}
\end{center}
\caption{Representative best antiproton fluxes for scenario I (left
  panel) and scenario II (right panel). Data refer to PAMELA
  measurement \cite{Adriani:2010rc} of the cosmic rays antiproton
  flux. The lower (blue) solid line refers to the theoretical
  determination of the secondary antiproton flux
  \cite{Donato:2003xg}. The dashed (blue) lines show a 10\%
  uncertainty on the secondary determination. The dotted (red) line
  shows the signal antiproton flux produced by dark matter
  annihilation: the best cases are considered, and they refer to
  $m_\chi = 51.1$ GeV for scenario I and to $m_\chi = 88.9$ GeV for
  scenario II. The upper (red) solid curve shows the sum of the
  secondary background and the signal.  In both panels, the MAX set of
  propagation parameters are used \cite{Donato:2003xg}.  }
\label{fig:pbar_spectra}
\end{figure*}

Current antiprotons bounds therefore do not constrain our
supersymmetric configuration, neither for the MIN nor for the MAX set
of propagation parameters. Prospects for future searches are shown in
Fig. \ref{fig:pbar028} by the dotted line, which refers to an expected
reach of AMS \cite{AMS}. We estimated AMS capabilities by taking into
consideration the following facts: AMS data on antiprotons will likely
reach a level of a few percent uncertainty; AMS will determine the
fluxes of cosmic rays species to an unprecedented level, and this will
help in reducing also the theoretical modeling of galactic cosmic rays
propagation. Determination of the boron--to--carbon (B/C) ratio will
be especially relevant. By considering a total (theoretical +
experimental) uncertainty on the antiproton fluxes after AMS, we can
estimate a bound (in case of non observation of deviation from the
expected background) at the level of the dotted lines in
Fig. \ref{fig:pbar028}: this would allow to probe a fraction of the
parameter space, both for scenario I and scenario II, in the case of
relatively large values of the propagation parameters (right panel of
Fig. \ref{fig:pbar028}). This capability is further illustrated in
Fig. \ref{fig:pbar_spectra}, where two representative antiproton
fluxes (one for scenario I and one for scenario II) are reported. The
two fluxes refer to the best--choice occurring in our parameter space,
but are nevertheless representative of those configurations with
fluxes in excess of the AMS reaching capabilities shown in the right
panel of Fig.  \ref{fig:pbar028}. Dark matter fluxes like those shown
in Fig. \ref{fig:pbar_spectra} will easily represent a detectable
signal in AMS, considering that they produce visible excesses over the
background (denoted by the solid line, while the dashed lines bracket
a $\pm$10\% uncertainty) in most of the energy spectrum.  We also
stress that AMS will have a very large statistics and therefore and
excess like those shown in Fig. \ref{fig:pbar_spectra} will be
detected as a deviation in a large number of experimental bins, thus
making the evidence of a signal potentially quite clear. The major
limitation remains the ability to reduce the theoretical uncertainties
on the background to a suitable level, as discussed above.

\begin{figure*}[t]
\begin{center}
\includegraphics[width=0.49\linewidth]{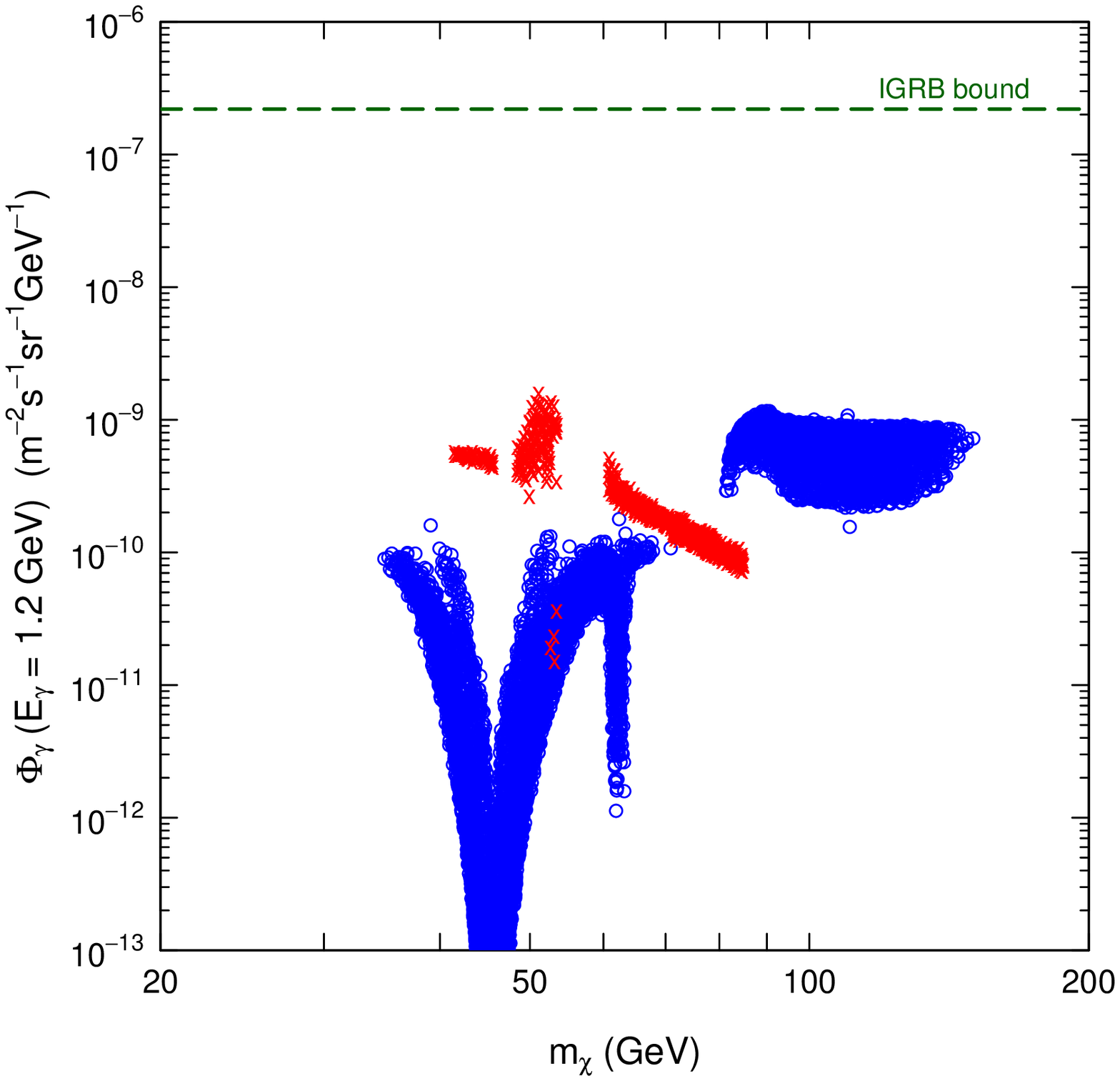}
\includegraphics[width=0.49\linewidth]{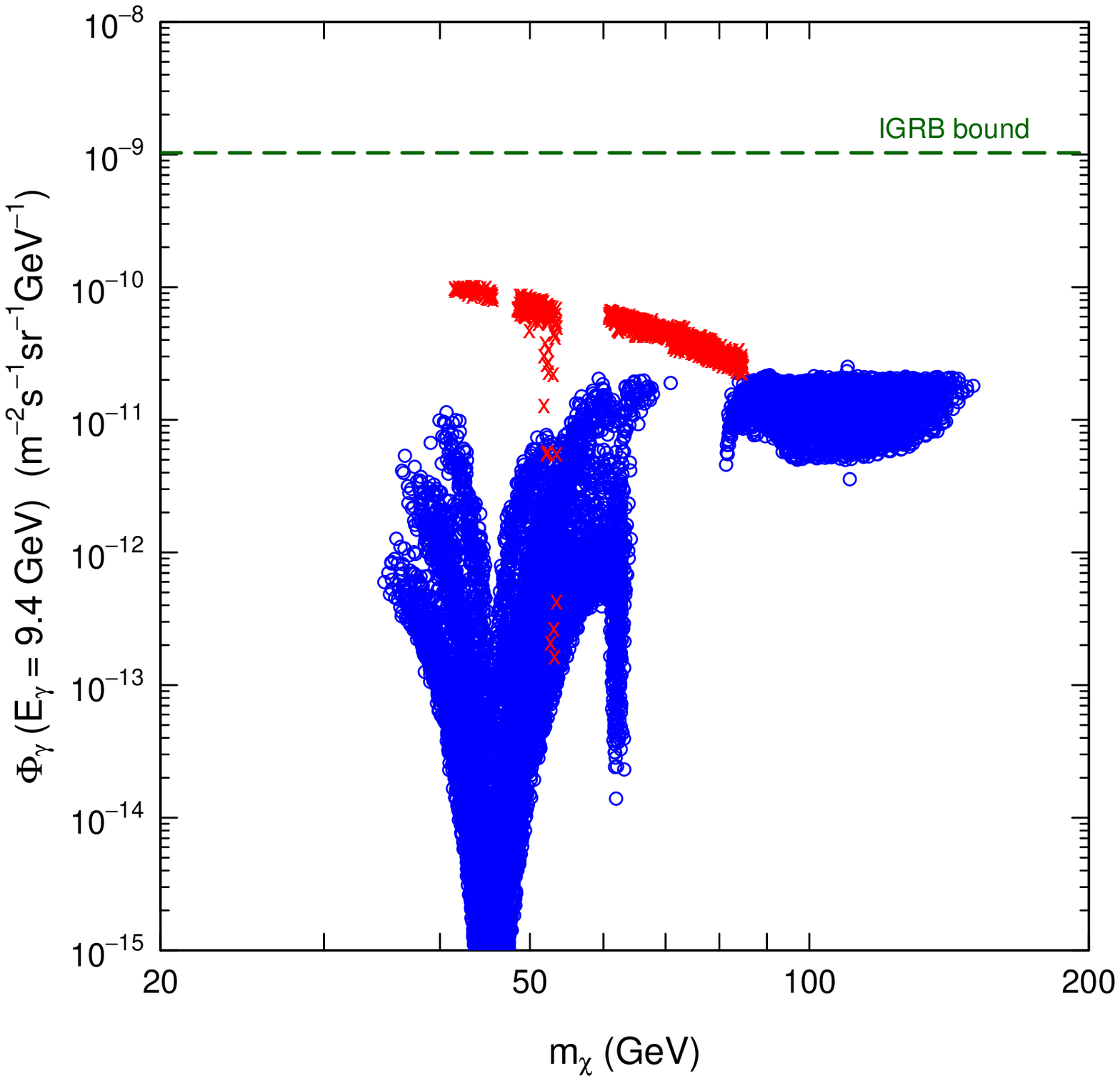}
\end{center}
\caption{Flux of gamma rays produced by galactic dark matter annihilation at high latitudes.
The horizontal dashed line represents the upper bound on the isotropic gamma--rays background
(IGRB) as determined by considering the Fermi--LAT \cite{Abdo:2010nz} 
measurements and the best--fit of various contributions to the IGRB \cite{Calore:2013yia}:
misaligned AGN \cite{DiMauro:2013xta}, star--forming galaxies \cite{Ackermann:2012vca}, 
unresolved milli--second pulsars \cite{Calore:2011bt}, BL Lacertae \cite{Collaboration:2010gqa} and flat-spectrum radio quasars \cite{Harding:2004hj}.
The two panels show the fluxes in two different energy bins of the Fermi--LAT analysis \cite{Abdo:2010nz}: the
left panel refers to $E_\gamma = 1.2$ GeV, the right panel to $E_\gamma = 9.4$ GeV.
(Red) crosses refer to configurations of scenario I; (blue) circles to scenario II.}
\label{fig:gamma}
\end{figure*}

Concerning the indirect signal in terms of gamma-rays,
Fig. \ref{fig:gamma} shows the flux of gamma rays produced by galactic
dark matter annihilation at high latitudes for both scenario I and
scenario II. The contribution to the IGRB has been calculated for an
Einasto profile of the dark matter density, but different profiles
predict only slightly different fluxes \cite{Calore:2013yia}, since we
are looking here at high galactic latitudes.

The signal fluxes in both scenario I and scenario II are relatively
small, when compared to the current upper bounds on the IGRB, obtained
by considering the Fermi--LAT measurements \cite{Abdo:2010nz} and the
best--fit of various contributions to the IGRB \cite{Calore:2013yia}:
misaligned AGN \cite{DiMauro:2013xta}, star--forming galaxies
\cite{Ackermann:2012vca}, unresolved milli--second pulsars
\cite{Calore:2011bt}, BL Lacertae \cite{Collaboration:2010gqa} and
flat-spectrum radio quasars \cite{Harding:2004hj}. The upper bound at
the 95\% C.L. is shown in Fig. \ref{fig:gamma} by the horizontal
dashed line. The figure shows the flux at two representative energies,
corresponding to two different energy bins of the Fermi--LAT analysis
\cite{Abdo:2010nz}: the left panel refers to $E_\gamma = 1.2$ GeV, the
right panel to $E_\gamma = 9.4$ GeV.

We notice that the contribution to the IGRB of astrophysical origin
suffers of large uncertainties: in deriving the bounds shown in
Fig. \ref{fig:gamma} we have adopted the central--value determinations
of the different sources of background, as reported in
Ref. \cite{Calore:2013yia}. If (just) some of these background fluxes
are allowed to fluctuate up (especially the recently determined
gamma-ray flux originating from misaligned AGN \cite{DiMauro:2013xta})
the ensuing bounds can become quite constraining
\cite{Calore:2013yia}.
%%%%%%%%%%%%%%%%%%%%%%%%%%%%%%%%%%%%%%%%%%%%%%%%%%%%%%%%%%%%%%%
%%%%%%%%%%%%%%%%%%%%%%%%%%%%%%%%%%%%%%%%%%%%%%%%%%%%%%%%%%%%%%%%

%%%%%%%%%%%%%%%%%%%%%%%%%%%%%%%%%%%%%%%%%%%%%%%%%%%%%%%%%%%%%%%
%%%%%%%%%%%%%%%%%%%%%%%%%%%%%%%%%%%%%%%%%%%%%%%%%%%%%%%%%%%%%%%%

%%%%%%%%%%%%%%%%%%%%%%%%%%%%%%%%%%%%%%%%%%%%%%%%%%%%%%%%%%%%%%%

%%%%%%%%%%%%%%%%%%%%%%%%%%%%%%%%%%%%%%%%%%%%%%%%%%%%%%%%%%%%%%%%

\section{Conclusions}
\label{sec:conclusions}

The attempt of interpreting the neutral boson ($H_{125}$) measured at the LHC in the diphoton, $Z Z$, $W W$ and $\tau \tau$ channels, and with a mass of
125--126 GeV, in terms of the effective Minimal Supersymmetric extension of the
Standard Model defined in Sect. \ref{sec:model}, has led us to consider
two possible scenarios: a scenario I, where the boson $H_{125}$ is identified
with the heavier CP--even neutral boson $H$ and scenario II,
where the boson $H_{125}$ is identified with the lighter CP--even neutral
boson $h$.

The supersymmetric parameter space has been analysed also in terms of
a full set of constraints derived from collider experiments, B--factories, and
measurements of the muon anomalous magnetic moment.
The properties of the neutralino as a dark matter constituent has been analysed
in both scenarios, considering its relic abundance and direct and indirect
detection rates.

We have found that in scenario I no solution for supersymmetric
configurations exists, unless two indirect constraints (BR($b
\rightarrow s + \gamma$) and $(g - 2)_{\mu}$) are relaxed. If these
two requirements are not implemented, solutions with a physical relic
abundance are found in a region of the supersymmetric parameter space
characterized by low values for the stau mass parameters $80 \, {\rm
  GeV} \leq m_{\tilde l_{12,L}}, m_{\tilde l_{12,R}}, m_{\tilde
  {\tau}_L}, m_{\tilde {\tau}_R} \leq 200 \, {\rm GeV }$, and high
values for the $\mu$ parameter: $\mu \geq$ 1.8 TeV. In the region
defined in Table \ref{tab:scenarioI} the neutralino mass turns out to
sit in the range $m_{\chi} \simeq (40-85)$ GeV.  The set of
configurations found in the present scenario generate very low rates
for direct detection of relic neutralinos (the quantity $\xi
\sigma_{\rm scalar}^{(\rm nucleon)}$ is at the level of $\xi
\sigma_{\rm scalar}^{(\rm nucleon)} \sim$ a few $\times 10^{-45}$
cm$^2$).  The same occurs for indirect detections signals: only
antiproton searches, under some optimistic assumptions, may be able to
test scenario I for neutralino masses close to 50 GeV. For this to be
reachable, a somehow large cosmic--rays confinement region is
required, accompanied by a reduction of the total theoretical +
experimental uncertainty on the antiproton flux determination at the
level of about 10\%. AMS \cite{AMS} is expected to beat this level of
precision on the antiproton data, and its measurement of the fluxes of
cosmic rays species, especially B/C, could help in reducing the
uncertainties on the theoretical determination, allowing to approach
the level required to study these supersymmetric populations.

In scenario II we have found a population of configurations which
satisfy all requirements and constraints mentioned in
Sect. \ref{sec:constraints}, including the indirect bounds coming from
BR($b \rightarrow s + \gamma$) and $(g - 2)_{\mu}$.  Here the lower
limit for the neutralino mass is $m_{\chi} \gsim$ 30 GeV. The direct
detection rates are shown to be typically rather low; though, they
could approach the level of the signals measured by the experiments of
DM direct detection \cite{dama2010,cresst,cogent,cdms2013} under special
instances for the DM distribution, for experimental parameters and/or
for significantly large size of the neutralino-nucleon coupling. As
for the indirect signals a situation similar to scenario I occurs:
under the same, somehow optimistic, assumptions discussed above an
antiproton signal in AMS may be reachable for neutralino masses above
80 GeV.

A few comments are in order here, regarding the features of the
population of relic neutralinos examined in the present paper: a) our
results apply only to the standard situation of thermal decoupling in
a standard FRW cosmology; in more extended cosmological scenarios,
especially those with an enhanced expansion rate of the Universe the
features of these populations are expected to be different
\cite{Catena:2004ba,Catena:2009tm,Gelmini:2009yh}; b) the relic
neutralinos considered here could constitute only a part of a
multicomponent DM (another component would be the one responsible for
the signals observed until now in DM direct detection experiments); c)
the derivations presented in the present paper rest heavily on the
results obtained at colliders: many of the analyses pertaining these
results are actually in progress, thus some of them could be
susceptible of significant modifications, with the implication of
possible substantial changes in the our present conclusions.

\acknowledgments
We thank F. Donato and M. Di Mauro for insightful discussions on the
IGRB and on its potential in bounding DM gamma--ray signals.  A.B. and
N.F. acknowledge the Research Grant funded by the Istituto Nazionale
di Fisica Nucleare within the {\sl Astroparticle Physics Project}
(INFN grant code: FA51).  S.S. acknowledges support by the National
Research Foundation of Korea(NRF) grant funded by the Korea
government(MEST) (No.2012-0008534). N.F. acknowledges support of the
spanish MICINN Consolider Ingenio 2010 Programme under grant MULTIDARK
CSD2009- 00064.

\end{document}